\documentclass[%
reprint,
superscriptaddress,
showpacs,preprintnumbers,
 amsmath,amssymb,
aps,
pre,
floatfix
]{revtex4-1}

\usepackage{float}
\usepackage{graphicx}
\usepackage{dcolumn}
\usepackage{bm}
\usepackage{bbold}
\usepackage[multidot]{grffile}
\usepackage[colorlinks=true,allcolors=blue, breaklinks=true]{hyperref}

\usepackage{xcolor}

\usepackage{xspace}

\usepackage[utf8]{inputenc}
\usepackage{tikz}
\usetikzlibrary{shapes.geometric,fit,arrows,babel}

\begin{document}


\title{Control of coherence resonance by self-induced stochastic resonance \\in a multiplex neural network}

\author{Marius E. Yamakou}
\email{yamakou@mis.mpg.de}
\affiliation{Max-Planck-Institut f\"{u}r Mathematik in den Naturwissenschaften, Inselstr. 22, 04103 Leipzig, Germany}
\author{J\"{u}rgen Jost}
\email{jost@mis.mpg.de}
\affiliation{Max-Planck-Institut f\"{u}r Mathematik in den Naturwissenschaften, Inselstr. 22, 04103 Leipzig, Germany}
\affiliation{Santa Fe Institute for the Sciences of Complexity, NM 87501, Santa Fe, USA}
%
%

\date{\today}

\begin{abstract}
We consider a two-layer multiplex network of diffusively coupled FitzHugh-Nagumo (FHN) neurons in the excitable regime. 
We show that the phenomenon of coherence resonance (CR) in one layer cannot only be controlled 
by the network topology, the intra and inter-layer time-delayed couplings, but also by another phenomenon, 
namely, self-induced stochastic resonance (SISR) in the other layer. Numerical computations show 
that when the layers are isolated, each of these noise-induced phenomena is weakened (strengthened)
by a sparser (denser) ring  network topology, stronger (weaker) 
intra-layer coupling forces, and longer (shorter) intra-layer time delays.  
However, CR shows a much higher sensitivity than SISR to changes in these control parameters.
It is also shown, in contrast to SISR in a single isolated FHN neuron, that the maximum noise amplitude at which 
SISR occurs in the network of coupled FHN neurons is controllable, especially in the regime of strong coupling forces and long time delays. 
In order to use SISR in the first layer of the multiplex network to control CR
in the second layer, we first choose the control parameters of the second layer in isolation
such that in one case CR is poor and in another case, non-existent. It is then shown that a pronounced SISR 
cannot only significantly improve a poor CR, but can also induce a pronounced CR, which was non-existent in the isolated second layer.
In contrast to strong intra-layer coupling forces, strong inter-layer coupling forces are found to enhance CR. 
While long inter-layer time delays just as long intra-layer time delays, deteriorates CR.
Most importantly, we find that in a strong inter-layer coupling regime, 
SISR in the first layer performs better than CR in enhancing CR in the second layer. 
But in a weak inter-layer coupling regime, CR in the first layer performs better than SISR in enhancing CR in the second layer.
Our results could find novel applications in noisy neural network dynamics and engineering.
\end{abstract}

\maketitle
\section{Introduction}
Noise is ubiquitous and even  relatively low noise intensities can have significant and
counter-intuitive effects on nonlinear dynamical systems. For instance,
a randomly perturbed system in the excitable regime can produce
oscillatory responses which possess a high degree of coherence and
yet are completely different from what is observed in the absence
of the random perturbations. In this paper, we focus on two such
phenomena: coherence resonance (CR) \cite{Hu et al 1993,Pikovsky
and Kurths 1997,Lindner 1999,Lindner 2004} and self-induced
stochastic resonance (SISR) \cite{Freidlin 2001,Freidlin2
2001,Muratov et al 2005} in a network of diffusively coupled FitzHugh-Nagumo (FHN)
neurons; a paradigmatic model which describes the excitability and
spiking behavior of neurons \cite{FitzHugh 1961}. Lee DeVille et
al. \cite{Lee DeVille et al. 2005} (see also \cite{Muratov et al
2008}) have analyzed  the differences in the mechanisms
leading to CR and SISR in an isolated excitable system. They
showed that even though CR and SISR lead to the emergence of the
same dynamical behavior (i.e., weak-noise-induced coherent
oscillations) in excitable systems, they are fundamentally 
different in their dynamical and emergent nature. 

CR occurs when the regularity of noise-induced oscillations of an
excitable system is a non-monotonic function of the noise
amplitude, and it is optimally correlated at some non-zero value of
the noise intensity. Thus, during CR, there exists a
maximal degree of coherence in the oscillations of the excitable
system for some intermediate noise amplitude. CR is known to be
robust in the sense that the coherence of the noise-induced
oscillations is insensitive against variations of the timescale
separation ratio between the fast and slow variables of the
excitable system and the amplitude of the noise \cite{Muratov et
al 2005,Lee DeVille et al. 2005}. The crucial condition
 necessary for the occurrence of CR is the proximity of the
system's parameters to the Hopf bifurcation \cite{Pikovsky and Kurths
1997,Neiman 1997,Lindner 1999,Lindner 2004,Beato 2007} or
the saddle-node bifurcation of limit cycles \cite{Hizanidis 2008,Liu
et al 2010,Jia et al 2011,Gu 2011}. This means that the system
must be near but before bifurcation thresholds so that a relatively small
noise amplitude can easily (without overwhelming the entire
dynamics) drive the system towards the deterministic limit cycle
which emerges right after the bifurcation. This implies that noise plays a rather 
passive role in the phenomenon of CR.

On the other hand, just like CR, SISR occurs when an optimal noise
amplitude induces coherent oscillations in a system in the
excitable regime (i.e., a regime that does not allow for
self-sustained oscillations in absence of perturbations). The similarity
between CR and SISR however ends at this point. Unlike CR, SISR is
robust to parameter tuning and therefore does not require the
system's parameter to be in the immediate neighborhood of bifurcation
thresholds. SISR relies completely on the ability of the slow-fast
excitable system to match the timescales of the motion of
trajectories along the slow manifold and  of escape
events of these trajectories from the slow manifold. The
characteristics of the coherent noise-induced oscillations due to
SISR have been shown to depend non-trivially on both the timescale
separation ratio weakly via its logarithm but more importantly on
the amplitude of the noise \cite{Freidlin 2001,Lee DeVille et al.
2005,Muratov et al 2008,Deville1 2007,Deville2 2007,Yamakou and Jost 2018,Yamakou and Jost 2017}.
This means that for SISR, noise plays a more subtle role in the
generation of the coherent oscillations than in CR.

For the specific case of the FHN model, CR occurs only when the
noise term is attached to the slow recovery variable equation (one
should note that this is not a necessary condition for the
occurrence of CR in general, as CR may still occur in slow-fast
systems where the noise term is attached to the fast variable
equation; see for example \cite{Beato 2007}). As pointed out
earlier, what is crucial for CR is the proximity of the system's
parameter to the bifurcation thresholds. Therefore, for CR, the noise term
has to be added to the system such that it can directly affect the
bifurcation parameter \cite{Hu et al 1993,Pikovsky and Kurths
1997,Lindner 1999,Lindner 2004}. This is why the noise term is
added to the slow variable equation of the FHN model in the case
of CR. 

On the other hand, because SISR occurs only in the singular
limit (i.e., the limit as the timescale separation ratio tends to
zero), it requires the noise term to be attached to the fast
variable equation of the slow-fast excitable system which evolves
itself on the fast timescale \cite{Freidlin 2001,Freidlin2
2001,Muratov et al 2005,Lee DeVille et al. 2005,Muratov et al
2008,Deville1 2007,Deville2 2007,Shen et al 2010,Yamakou and Jost 2018}.
Because SISR is more stable than CR to parametric perturbations, 
it might be useful from an engineering point of view to try to control CR using SISR 
and not the other way round.

One of the most relevant questions today is related to the control
of noise-induced phenomena in networks of coupled
oscillators. Several control schemes of CR based on time-delayed
feedbacks and network topology and heterogeneity have been
studied, in particular with the FHN model equations. It has been shown that
appropriate selection of the time-delayed feedback parameters can
modulate CR \cite{Janson et al 2004,Hizanidis et al 2006,Hizanidis
2008,Aust et al 2010,Geffert et al 2014,Semenov et al 2015,Maria
et al 2017}. With a particular version of the FHN model, it has been shown that in a ring of locally coupled 
units, time-delay weakens CR while in non-local and global
coupling cases, only appropriate delay values could strengthen or
weaken CR \cite{Maria et al 2017}. C. Zhou et at \cite{Zhou et al
2001} showed that the spatial heterogeneity of the bifurcation
parameter of coupled FHN oscillators can also control CR. In this
control scheme, CR is enhanced in the network producing
oscillations that are more coherent than those generated during CR
in a single FHN unit. It has also been shown that in a ring of non-locally coupled 
oscillators, time-delays can control the parameters range in which CR-induced chimera states occur \cite{Zakharova et al 2017}. 

Recently, the use of multiplex network in the control strategy of 
dynamical behaviors of coupled oscillators has attracted some attention.
In a multiplex network, each type of interaction between the nodes
is described by a single layer network and the different layers of
networks describe the different modes of interaction. In such
networks, the layers contain the same number of nodes and the
interaction between the layers are allowed only for replica nodes.
In the case of neural networks,
the neurons can form different layers depending on their
connectivity through a chemical link or by an ionic channel. In
brain networks, different regions can be seen connected by
functional and structural neural networks \cite{Domenico
2017,Pisarchik 2014,Andreev et al 2018}.

Multiplex networks open up new possibilities of control, allowing
to regulate nonlinear systems by means of the interplay between
dynamics and multiplexing. Control mechanisms based on
multiplexing have many advantages. In particular, the dynamics of one layer can be controlled by adjusting the parameters
of another layer. This is important from the point of view of
engineering and brain surgery since it is not always possible to
directly access the desired layer, while the network with which
this layer is multiplexed may be accessible and adaptable. The
multiplexing of networks has been shown to control many dynamical
behaviors including synchronization \cite{Gambuzza et al
2015,Singh et al 2015,Leyva et al 2017,Zhang et al 2017,Andrzejak
et al 2017} and pattern formation \cite{Kouvaris et al 2015,Ghost
and Jalan 2016,Ghosh et al 2016,Maksimenko et al 2016,Bera et al
2017,Bukh et al 2017,Ghosh et al 2018}. However, the control of CR
based on the multiplexing of networks has not been extensively
investigated. This alternative control scheme of CR has been only
very recently studied in \cite{Semenova1 and Zakharova 2018}. Here, it
is shown that even weak multiplexing can induce CR in a network of FHN neurons that does not
demonstrate CR in isolation. Moreover, it has been shown that the multiplex-induced CR in
the layer which is deterministic in isolation can manifest itself
even more strongly than that in the noisy layer. In \cite{Semenova1 and Zakharova 2018}, 
each layer of the multiplex network considered undergoes CR, as the noise term in attached only
to the slow variable equations and the bifurcation parameter is set in the non-oscillatory regime 
but close to the Hopf bifurcation threshold.
  
The objective of this paper is to propose a new control strategy of
CR in a two-layer multiplex network of FHN neurons in the excitable regime.
Our control strategy is not only based on the combined effects of time-delayed couplings, intra-layer network topology, and multiplexing, 
but will also include the phenomenon of SISR. In short, we aim at controlling CR 
with SISR in a time-delayed multiplex network. Our aim and setting are 
different from the one in \cite{Semenova1 and Zakharova 2018} which considers the control of CR in one layer of the multiplex 
with CR in the other layer and which ignores time-delays (which are ubiquitous in complex systems). 
 The main questions we want to address in this work are the following:\\ 
(i) In a two-layer multiplex network, can SISR occurring in one layer be used to control CR in the other layer? \\
(ii) Which phenomenon (i.e., SISR or CR, occurring in one layer) would better enhance CR in the other layer?

This paper is organized as follows: In section \ref{section2}, we present the model equation and determine the parameter range
of its excitable regime. In section \ref{section3}, we present our results. In the first part of this 
section, we consider isolated layers and investigate the effects of time-delayed couplings and different ring network topologies on 
CR and SISR. In the second part, we consider the multiplex network of the layers where we present and compare the control schemes of CR based on SISR and on CR. 
In section \ref{section4}, we have the summary and conclusion.
 
\section{Model and its excitable regime}\label{section2}
\begin{figure}
\begin{center}
  \begin{tikzpicture}[baseline=(current  bounding box.center),>=stealth]
   \node[inner sep=0pt] (fig1) at (0,0)
    {\includegraphics[scale=0.45,keepaspectratio=true]{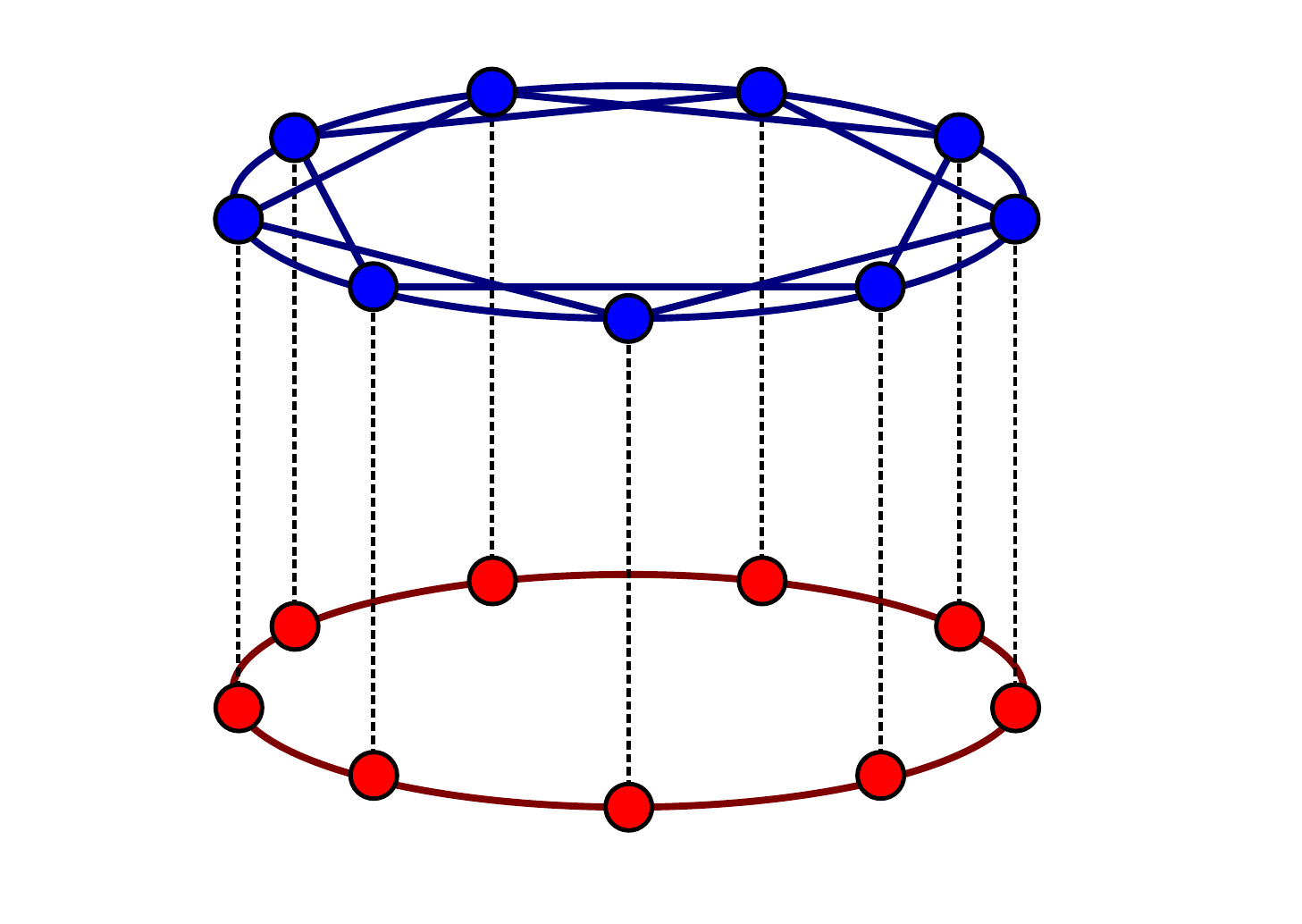}};
    \node at (3.25,1.2) {layer 1: $\kappa_{_1},\tau_{_1}$};
     \node at (3.5,0.0) {multiplexing: $\kappa_{_{12}},\tau_{_{12}}$};
    \node at (3.25,-1.2) {layer 2: $\kappa_{_2},\tau_{_2}$};
  \end{tikzpicture}
  \caption{(Color online) Schematic diagram showing a multiplex network of neurons consisting of two layers. Each node in a layer
  is connected only to the replica node in the other layer.
  The lower layer (layer 2) in red represents a coupled ring network topology, 
  where CR occurs. The upper layer (layer 1) in blue represents a coupled ring network (with possibly a different ring topology), 
  where either SISR or CR can occur and be used together with the 
  intra and inter time-delayed couplings parameters $\{\tau_{_1}, \kappa_{_1}, \tau_{_2}, \kappa_{_2}, \tau_{_{12}}, \kappa_{_{12}}\}$, to control CR in layer 2.}\label{fig:1a}
\end{center}
\end{figure}
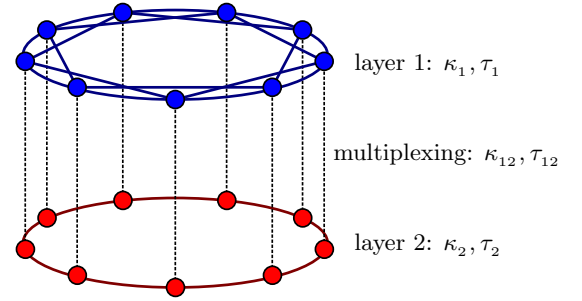

We consider the following two-layer multiplex network, where each layer
represents a ring of $N$ diffusively coupled FHN neurons in the excitable regime and in the presence of noise:
\begin{eqnarray}\label{eq:1}
\begin{split}
\left\{\begin{array}{lcl}
\displaystyle{\frac{dv_{_{1i}}}{dt}}&=&v_{_{1i}}-\displaystyle{\frac{v_{_{1i}}^3}{3}}-w_{_{1i}}\\[3.0mm]
&+&\frac{\kappa_{_1}}{2n_{_1}}\sum\limits_{j=i-n_{_1}}^{i+n_{_1}}\Big(v_{_{1j}}(t-\tau_{_1})-v_{_{1i}}(t)\Big)\\[3.5mm]
&+&\kappa_{_{12}}\Big(v_{_{2i}}(t-\tau_{_{12}})-v_{_{1i}}(t)\Big) +\sigma_{_1}\frac{dW_{_{1i}}}{dt},\\[1.0mm]
\displaystyle{\frac{dw_{_{1i}}}{dt}}&=&\varepsilon_{_1}(v_{_{1i}}+\alpha-\beta_{_1} w_{_{1i}})+\sigma_{_3}\frac{dW_{3i}}{dt},\\[3.0mm]
\displaystyle{\frac{dv_{_{2i}}}{dt}}&=&v_{_{2i}}-\displaystyle{\frac{v_{_{2i}}^3}{3}}-w_{_{2i}}\\[3.0mm]
&+&\frac{\kappa_{_2}}{2n_{_2}}\sum\limits_{j=i-n_{_2}}^{i+n_{_2}}\Big(v_{_{2j}}(t-\tau_{_2})-v_{_{2i}}(t)\Big)\\[3.5mm]
&+&\kappa_{_{12}}\Big(v_{_{1i}}(t-\tau_{_{12}})-v_{_{2i}}(t)\Big),\\[1.0mm]
\displaystyle{\frac{dw_{_{2i}}}{dt}}&=&\varepsilon_{_2}(v_{_{2i}}+\alpha-\beta_{_2} w_{_{2i}})+\sigma_{_2}\frac{dW_{_{2i}}}{dt}.
\end{array}\right.
\end{split}
\end{eqnarray}
$v_{_{1i}}\in\mathbb{R}$ and $w_{_{1i}}\in\mathbb{R}$ represent the fast membrane potential variables and the slow 
recovery current variables in the first layer, respectively. The index $i=1,...,N$ stands for the node $i$ in the ring network of $N$ neurons
and all indices are modulo $N$.
Similarly, $v_{_{2i}}\in\mathbb{R}$ and $w_{_{2i}}\in\mathbb{R}$ respectively
represent the membrane potential and recovery current variables for the neurons in the second layer.
$0<\varepsilon_{_1}\ll1$ and $0<\varepsilon_{_2}\ll1$ are the time-scale separation ratios between the fast membrane potential and the slow recovery current variables 
in the first and second layers, respectively. $\beta_{_1}>0$ and $\beta_{_2}>0$ are co-dimension-one Hopf bifurcation parameters in each of the layers and thus define the excitability 
threshold. $\alpha\in(0,1)$ is a constant parameter. For the $i^{th}$ neuron in a ring, $n_{_1}$ and $n_{_2}$ represent the number of 
nearest connected neighbors in each direction on the ring. In each ring, we can have 3 different coupling schemes between the neurons:
$n_{_{1,2}}=1$ for local coupling, $n_{_{1,2}}=(N-1)/2$ (for odd $N$) for global coupling, and $1<n_{_{1,2}}<(N-1)/2$ for non-local coupling. 
Hence, $n_{_1}$ and $n_{_2}$ acts as independent control parameters for the ring topology of the underlying networks of the multiplex.
The coupling strength within the layers (intra-layer coupling) is measured by $\kappa_{_1}$ and $\kappa_{_2}$ in the first and second layer, respectively. 
The intra-layer couplings have the form of a diffusive coupling, that is in each layer, the coupling term vanishes if $v_{_{1i}}$ and $v_{_{1j}}$ (resp. $v_{_{2i}}$ and $v_{_{2j}}$) 
are equal. $\tau_{_1}$ and $\tau_{_2}$ are respectively the signal propagation delays within the first 
and second layer, which we henceforth refer to as layer 1 and layer 2, respectively. The coupling between the layers 
(inter-layer coupling) is bidirectional, diffusive, and its strength is measured by $\kappa_{_{12}}$. $\tau_{_{12}}$ 
is the signal propagation delay between layer 1 and layer 2 of the multiplex network (see Fig.\ref{fig:1a}). $\sigma_{_1}$, $\sigma_{_2}$, and $\sigma_{_3}$ stand for noise intensities.
$\frac{dW_{_{ki}}}{dt}$, $k=1,2,3$ are uncorrelated Gaussian white noises, the formal derivative of Brownian motion with mean zero and unit variance.

The conditions required for the occurrence of CR and SISR  in an isolated FHN neuron are well known
\cite{Pikovsky and Kurths 1997,Lindner 1999,Muratov et al 2005,Lee DeVille et al. 2005,Yamakou and Jost 2018}. 
An excitable regime is a necessary and common requirement for both CR and SISR. 
In this regime, the isolated FHN neuron has a unique and stable fixed point. 
Choosing an initial condition in the basin of attraction of this fixed point will result in at
most one large non-monotonic excursion into the phase
space after which the trajectory returns to this fixed point and stays there until the initial
conditions are changed again \cite{Izhikevich 2000,Yamakou and Jost 2018}. 

In any isolated layer ($\kappa_{_{12}}=0$) of the multiplex network, we can set an isolated neuron ($\kappa_{_1}=0$ or $\kappa_{_2}=0$) in an excitable regime 
by fixing the parameters to certain values: We set $\alpha=0.5$ in each layer network. The bifurcation parameters $\beta_{_1}$ and $\beta_{_2}$ are chosen such that 
$\beta_{_1}>\beta_{_h}(\varepsilon_{_1})$ and $\beta_{_2}>\beta_{_h}(\varepsilon_{_2})$, where $\beta_{_h}(\varepsilon_{_1})$ and
$\beta_{_h}(\varepsilon_{_2})$ are the Hopf bifurcations values of a neuron in layer 1 and layer 2, respectively.
For each neuron in layer 1 and in layer 2, we set $\varepsilon_{_1}=0.0005$ and $\varepsilon_{_2}=0.01$ and calculate the Hopf 
bifurcation values as $\beta_{_h}(\varepsilon_{_1})=0.7497$ and $\beta_{_h}(\varepsilon_{_2})=0.7446$, respectively. 
It is worth noting that for $\beta_{_1}\leq\beta_{_h}(\varepsilon_{_1})$ and $\beta_{_2}\leq\beta_{_h}(\varepsilon_{_2})$, an isolated neuron is in the oscillatory regime; 
a regime that we want to avoid as the coherent oscillations generated by SISR are 
due only to the presence of noise and not because of the occurrence of a Hopf bifurcation \cite{Yamakou and Jost 2018}.
While the coherent oscillations generated by CR, even though they require proximity to the Hopf bifurcation threshold, 
they still need the system to be in the excitable regime \cite{Pikovsky and Kurths 1997}. 

To investigate the effects of time-delayed coupling and different ring network topologies on SISR in layer 1 and CR in layer 2, 
we have to make sure that for each layer, the entire network is in the excitable regime and not just the individual neurons.
Therefore, each layer should be in the parameter regime where there are no time-delayed coupling induced oscillations. 
It is important to identify this regime (if it exists) because certain time-delayed couplings may induce 
self-sustained oscillations in a layer network, even when all of the individual neurons are in the excitable regime.
A saddle-node bifurcation resulting in a pair of stable and unstable limit cycles may generate these delayed-coupling induced 
self-sustained oscillations \cite{Schoell et al. 2009}. 

To ensure that the deterministic isolated layers are in a parameter regime where time-delayed coupling induced oscillations are absent, 
we numerically calculate the oscillation period, i.e., the interspike interval ($ISI$) of the synchronized oscillations 
for different values of $\tau_{_1}$ and $\kappa_{_1}$ in layer 1 ($\varepsilon_{_1}=0.01$) and $\tau_{_2}$ and $\kappa_{_2}$ in layer 2 ($\varepsilon_{_2}=0.0005$).
For the values of $\beta_{_1}$ and $\beta_{_2}$ used in the calculations, layer 1 and layer 2 showed the same behavior. 
We therefore only show the results of layer 2. Throughout this paper, we will for the purpose of 
simplicity, set $\beta_{_1}=\beta_{_2}=0.75$. At this value, each layer is in the excitable regime. 

\begin{figure}
\begin{center}
\includegraphics[width=4.25cm,height=3.6cm]{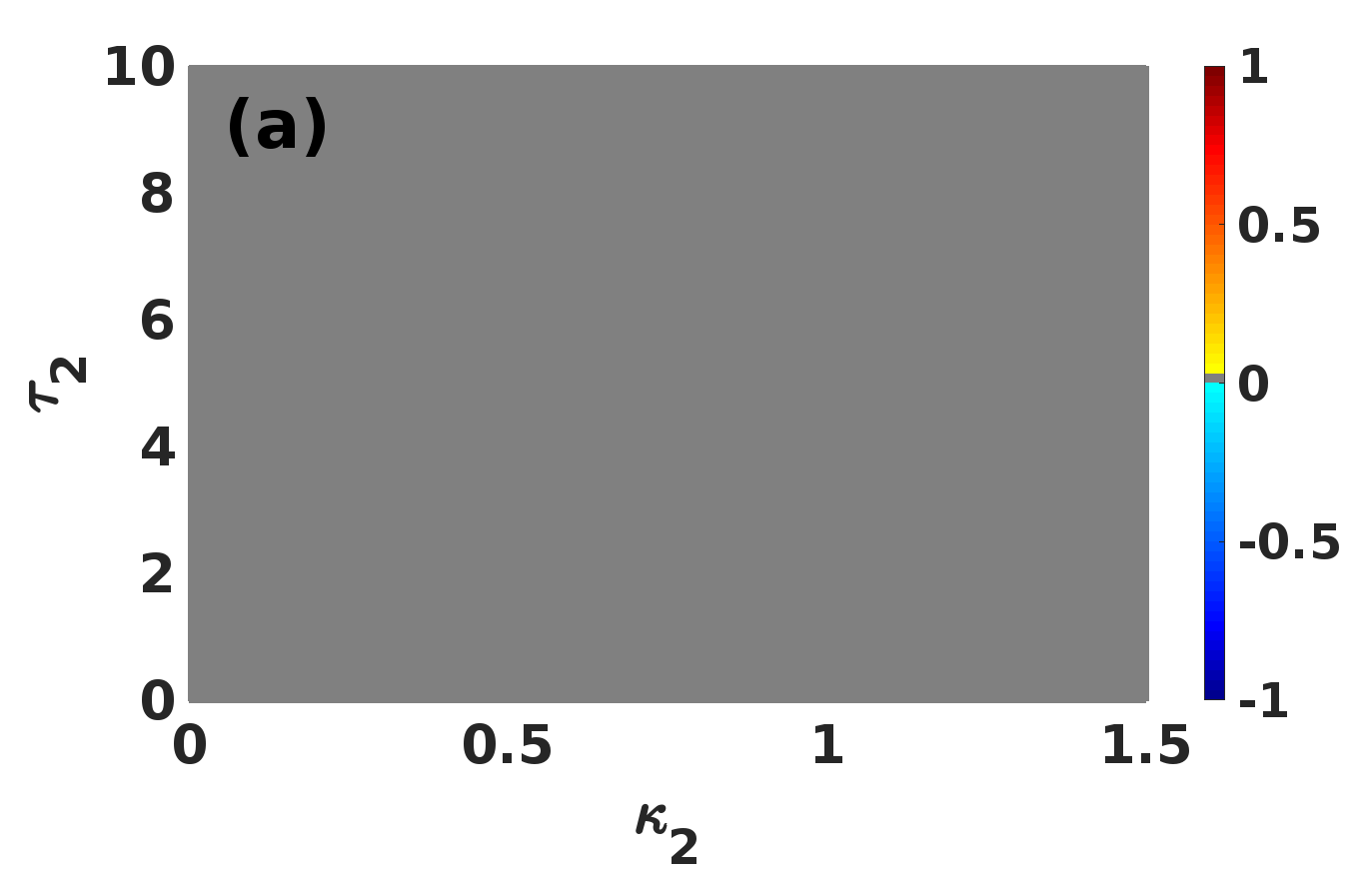}\includegraphics[width=4.25cm,height=3.6cm]{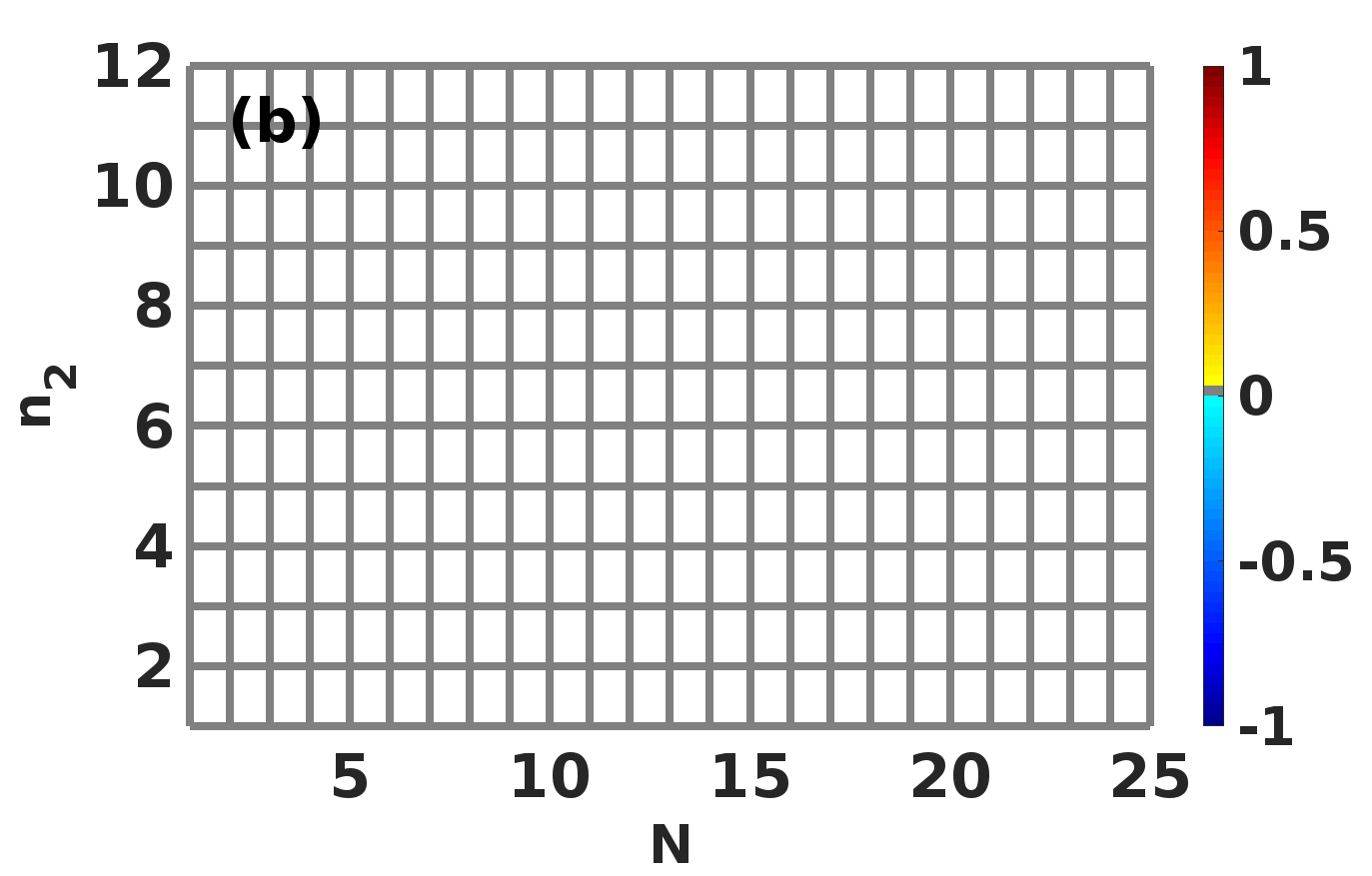}
\includegraphics[width=4.25cm,height=3.6cm]{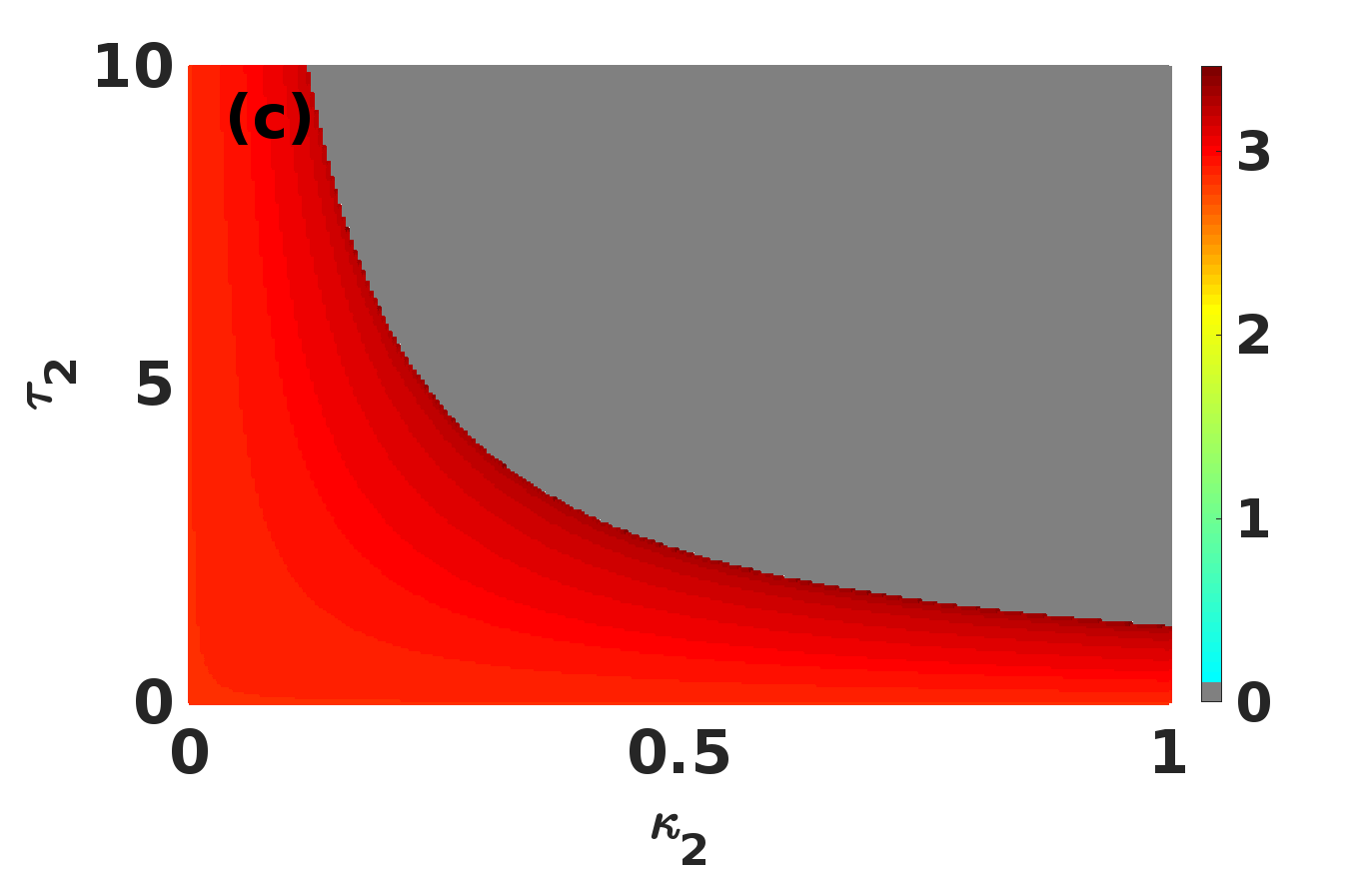}\includegraphics[width=4.25cm,height=3.6cm]{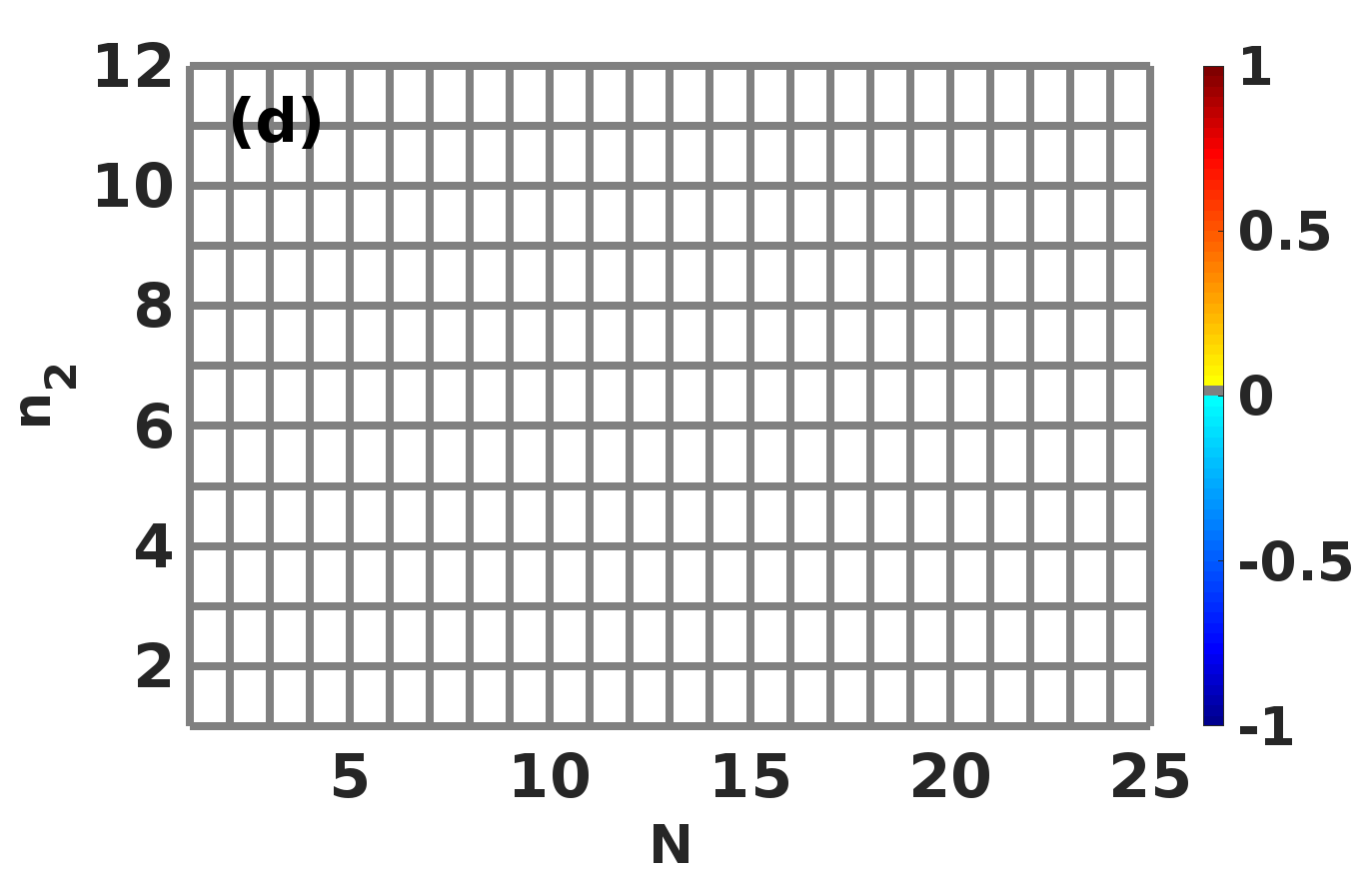}
\includegraphics[width=4.25cm,height=3.6cm]{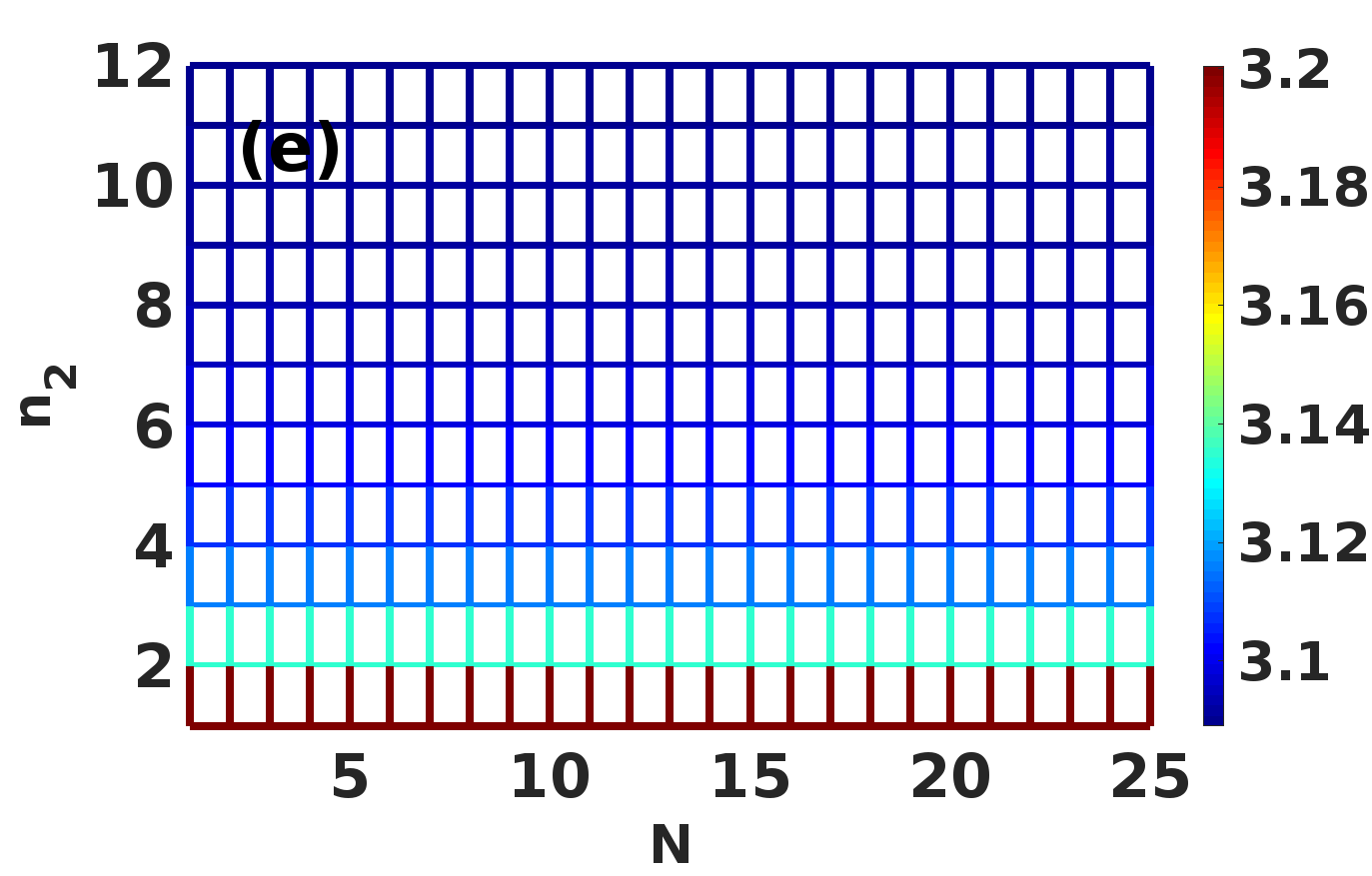}
\caption{(Color online) The $ISI$ is color coded and corresponds to $ISI + \xi$ with $0<\xi\ll1$. 
\textbf{(a):} the gray region in the $(\kappa_{_2}-\tau_{_2})$ plane represents the excitable regime (consisting of a stable homogeneous fixed point) 
of the isolated layer 2 with the Hopf bifurcation parameter $\beta_{_2}=0.7500>\beta_{_h}(\varepsilon_{_2})=0.7446$, $n_{_2}=1$. \textbf{(b):} the excitable regime $(\kappa_{_2},\tau_{_2})=(0.5,5.0)$ in \textbf{(a)} is independent of $N$ and $n_{_2}$.
\textbf{(c):} the red region represents the oscillatory regime $\beta_{_2}=0.7400<\beta_{_h}(\varepsilon_{_2})$ and the gray region 
an excitable regime induced by amplitude death, $n_{_2}=1$. \textbf{(d):} the excitable regime 
$(\kappa_{_2},\tau_{_2})=(0.75,7.0)$ in \textbf{(c)} is independent of 
$N$ and $n_{_2}$. \textbf{(e):} the oscillatory regime $(\kappa_{_2},\tau_{_2})=(0.25,2.0)$ 
in \textbf{(c)} is independent of $N$ and $n_{_2}$. Initial conditions corresponds to a spike for all neurons. 
Other parameters of the isolated layer 2: $N=25$, $\varepsilon_{_2}=0.01$, $\alpha=0.5$, $\sigma_{_2}=0.0$.\label{fig:2}}
\end{center} 
\end{figure}

In Fig.\ref{fig:2}, $ISI$ is plotted in the $(\kappa_{_2}-\tau_{_2})$ and $(N-n_{_2})$ parameter spaces for different values of the Hopf bifurcation parameter $\beta_{_2}$. 
In Fig.\ref{fig:2}\textbf{(a)} and \textbf{(b)}, $\beta_{_2}>\beta_{_h}(\varepsilon_{_2})$.
While in Fig.\ref{fig:2}\textbf{(c)} and \textbf{(d)},  $\beta_{_2}\leq\beta_{_h}(\varepsilon_{_2})$.
We observe that there are no time-delayed coupling oscillations in Fig.\ref{fig:2}\textbf{(a)} and \textbf{(b)}. In this case, 
the excitable regime of each neuron persists in the entire layer network for the considered delayed-coupling intervals, network size $N$, 
and number of neighbors $n_{_2}$.
The gray region in Fig.\ref{fig:2}\textbf{(a)} and gray lines in Fig.\ref{fig:2}\textbf{(b)} represent the absence of
\textit{time-delayed coupling induced oscillations}: 
this is the excitable regime in which we focus our study and which consists of a stable homogeneous 
steady state given by $(v^*_{_{2i}},w^*_{_{2i}})=(-1.003975,-0.666651)$.

In contrast, in Fig.\ref{fig:2}\textbf{(c)} and \textbf{(d)}, 
when all the neurons in the layer are in the oscillatory regime (i.e., $\beta_{_2}\leq\beta_{_h}(\varepsilon_{_2})$), 
we instead have the phenomenon of \textit{amplitude death} which is characterized 
by oscillation quenching induced by the presence of delay in the coupling between the 
neurons in the network \cite{Reddy et al 1998,Reddy et al 2000,Resmi et al. 2001,koseska et al 2013}. Fig.\ref{fig:2}\textbf{(b)} and \textbf{(d)} show 
that the excitable regime of the layer network is independent of its size $N$ and the number of connected neighbors $n_{_2}$.
The oscillatory regime in Fig.\ref{fig:2}\textbf{(c)} (red region) also shows (in Fig.\ref{fig:2} \textbf{(e)}) an independent behavior with respect to $N$ and $n_{_2}$,
except that the period of the oscillations increases with decreasing number of connected neighbors $n_{_2}$.

\section{Methods and Results} \label{section3}
The coefficient of variation ($CV$), the correlation time, the power spectral density, and the signal-to-noise ratio are commonly 
used measures to qualify and quantify the regularity of noise induced oscillations,
and therefore a measure of how pronounced CR and SISR can be at a particular noise amplitude. 
From a neurobiological point of view, $CV$ is more important because 
it is related to the timing precision of the information processing in neural systems \cite{Pei et al 1996}. 
For a Poissonian spike train (rare and incoherent spiking), $CV=1$.
If $CV<1$, the sequence becomes more coherent, and $CV$ vanishes for a periodic deterministic spike train. 
$CV$ values greater than $1$ correspond to a point process that is more variable than a Poisson process \cite{Yamakou and Jost 2018,Kurrer and Schulten 1995}.
Because of the importance of $CV$ in neural information processing, 
we shall use it to characterize the regularity of the noise-induced oscillations generated by SISR and CR in our neural network. 
The $CV$ of an isolated neuron is defined as \cite{Pikovsky and Kurths 1997}:
\begin{equation}\label{eq:2}
CV=\frac{\sqrt{\langle ISI^2\rangle-\langle ISI\rangle^2}}{\langle
ISI\rangle},
\end{equation}
where $\langle ISI\rangle$ and $\langle ISI^2\rangle$ represent
the mean and the mean squared interspike intervals, respectively.
The above definition of $CV$ is limited to characterizing CR and SISR in an isolated neuron.
For a network of coupled neurons, CR and SISR can be measured by redefining $CV$ as follows \cite{Masoliver et al 2017}:
\begin{equation}\label{eq:48}
CV=\frac{\sqrt{\langle\overline{ISI^2}\rangle-\langle\overline{ISI}\rangle^2}}{\langle\overline{ISI}\rangle},
\end{equation}
where the extra bar indicates the additional average over neurons in the layer network.

In the rest of our numerical calculations, we use the fourth-order Runge-Kutta algorithm for
stochastic processes \cite{Klasdin 1995} to integrate over a very long time interval ($600000$ time
units) and then averaging over time, neurons ($N=25$), and $5$ realizations of each noise amplitude. In the numerical simulations, this long time interval 
permitted us to collect with a small noise amplitude at least $70500$ ISIs with $\varepsilon_{_2}=0.01$ used for CR 
and at least $125$ ISIs with $\varepsilon_{_1}=0.0005\ll\varepsilon_{_2}$ used for SISR.

\subsection{Isolated layers ($\kappa_{_{12}}=0$): Network topology and time-delayed coupling effects on CR and SISR}
In order to control (strengthen or weaken) CR in layer 2 of the multiplex network using SISR (or CR itself) in layer 1, we should naturally first 
understand the effects of the intra-time-delayed couplings and different ring network topologies on 
CR and SISR in the isolated  layers. We therefore begin our study with the dynamics of disconnected  ($\kappa_{_{12}}=0$)
layers and consider an isolated one-layer network. 

\subsubsection{\textbf{Coherence Resonance}}
The conditions required for the occurrence of CR in an isolated ($\kappa_{_2}=0$) FHN neuron in layer 2 are: the proximity of the
parameter to the Hopf bifurcation value 
and a sufficiently small noise amplitude that does not overwhelm the entire dynamics, and which is also sufficiently large 
so that trajectories do not stick for too long in the basin of attraction of the stable fixed point to produce a Poisson sequence of spikes
\cite{Pikovsky and Kurths 1997,Muratov et al 2005,Lee DeVille et al. 2005,Muratov et al 2008}, that is:
\begin{eqnarray}\label{eq:3}
\begin{split}
\left\{\begin{array}{lcl}
\beta_{_2}-\beta_{_h}(\varepsilon_{_2})\leq\delta,\\
\sigma_{_2}\ll1,\\
\sigma_{_2}\gtrsim(1-w_{_{2i}}^{*2}-\varepsilon_{_2}\beta_{_2})^{3/2}.
\end{array}\right.
\end{split}
\end{eqnarray}
where $\delta >0$ is small. We notice that in layer 2, the noise term with amplitude $\sigma_{_2}$ is attached only to the slow variable equations.

It has been shown that CR can be enhanced in a network of coupled oscillators by choosing 
appropriate coupling strengths and time delays \cite{Masoliver et al 2017,
Bambi Hu and Changsong Zhou 2000}. 
But one should note that the response of CR to changes in time-delayed coupling can vary from one kind of 
nonlinear oscillator to another depending on the details of the equation. 
That is, there is no general behavior as stronger coupling forces may strengthen CR in a network of one kind of 
nonlinear oscillators \cite{Semenova1 and Zakharova 2018}, but as we shall see in the current paper, 
it can instead weaken CR in a network of just slightly modified equations of the nonlinear oscillators.
Therefore, without investigations, one cannot say a \textit{priori} how CR will behave in response to changes 
in time-delayed couplings in a network of a particular nonlinear oscillator,
even if this behavior is known for a network of another nonlinear oscillator, with a slightly modified equation.

\begin{figure}
\begin{center}
\includegraphics[width=4.5cm,height=4.25cm]{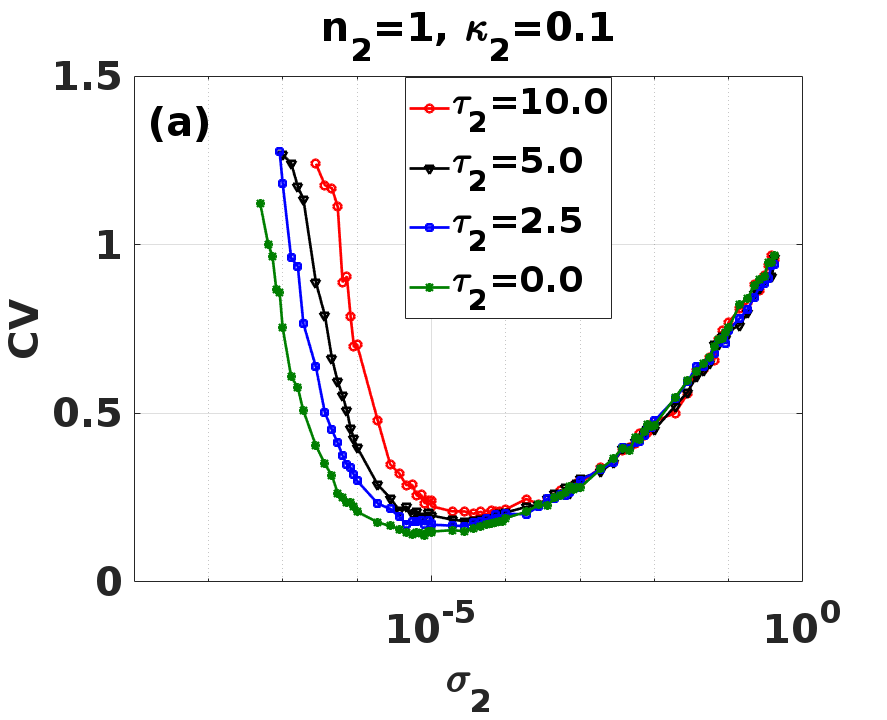}\includegraphics[width=4.5cm,height=4.25cm]{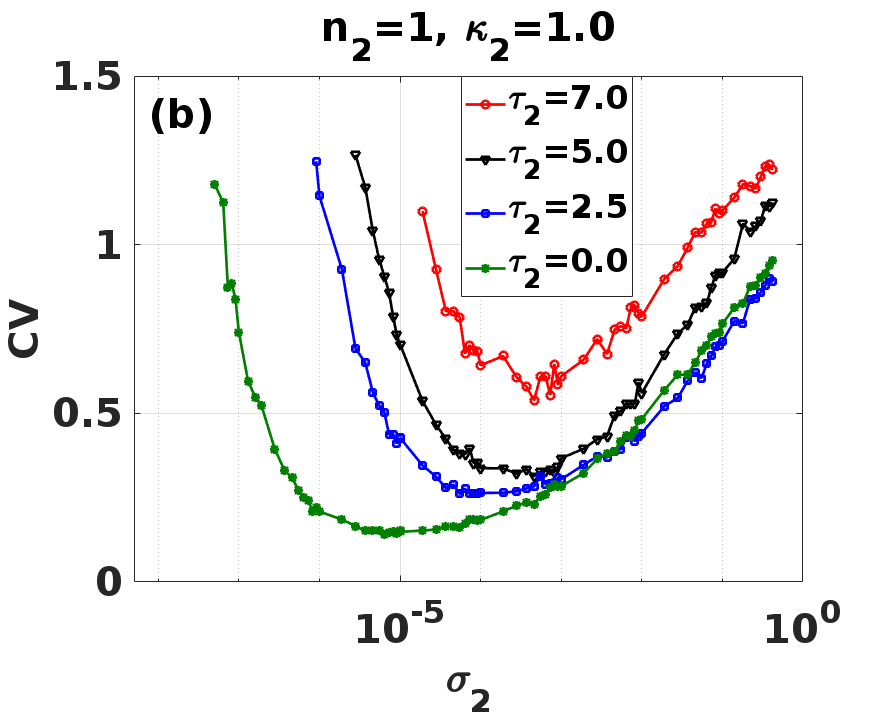}
\includegraphics[width=4.5cm,height=4.25cm]{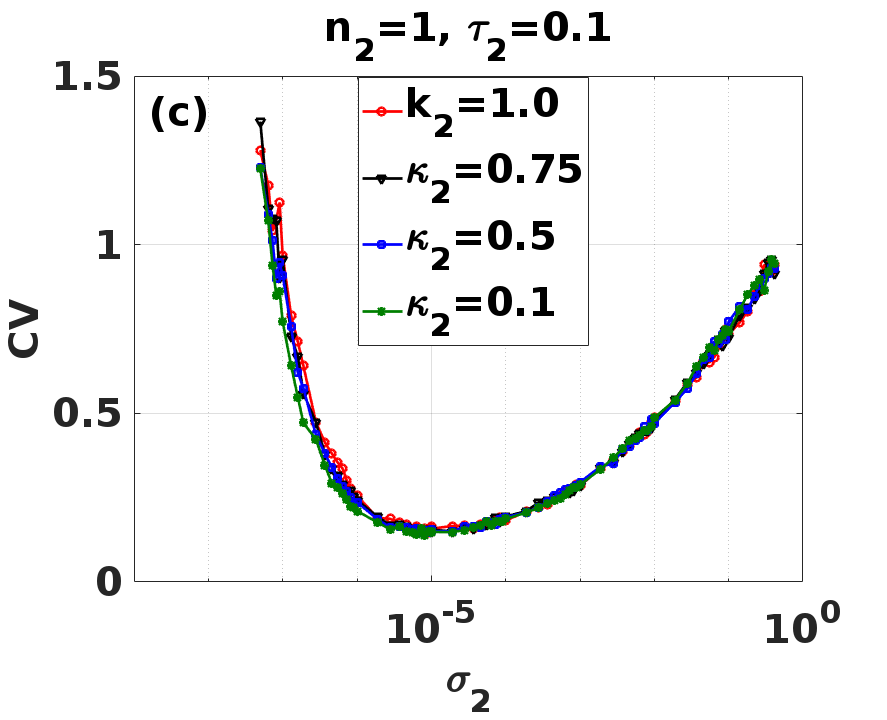}\includegraphics[width=4.5cm,height=4.25cm]{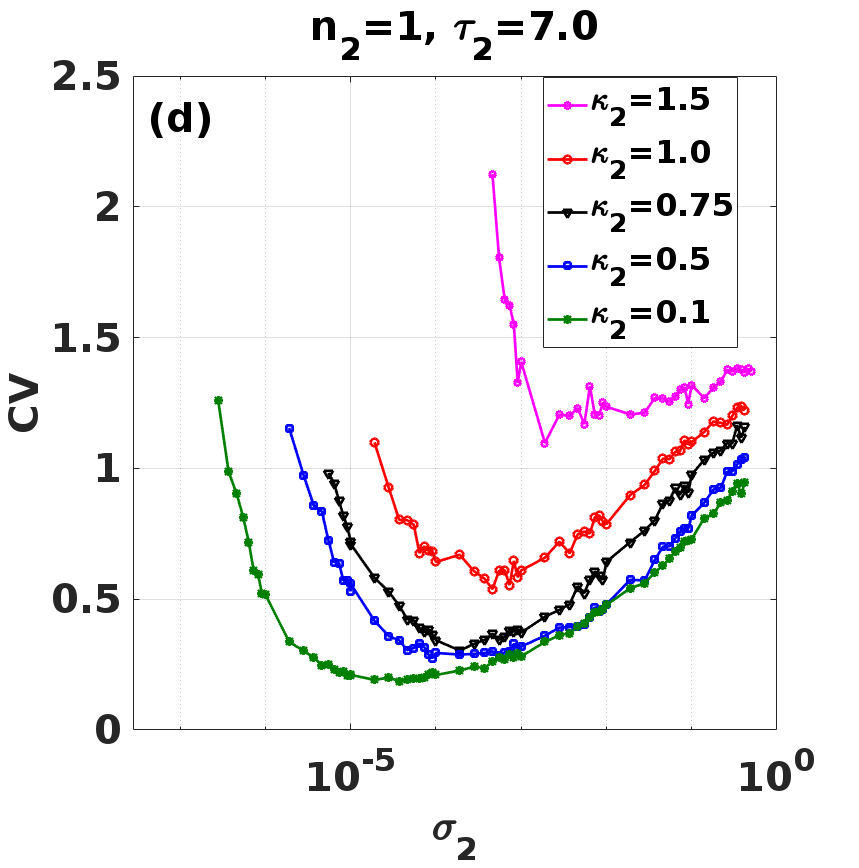}
\caption{(Color online) Coefficient of variation $CV$ against noise amplitude $\sigma_{_2}$ of layer 2 in isolation.
Increasing the strength of coupling force and the length of time delay shift the $CV$-curves up and to the right, thus weakening CR. 
In particular, in \textbf{(b)}, for $\kappa_{_2}=1.0$ and $\tau_{_2}=7.0$, the red  $CV$-curve has a minimum of $CV_{min}=0.56$ at $\sigma_{_2}=4.6\times10^{-4}$ indicating 
a poor CR. In \textbf{(d)}, for $\kappa_{_2}=1.5$ and $\tau_{_2}=7.0$, the pink $CV$-curve lies entirely above the line $CV=1.0$, indicating the absence
of CR. Parameters of layer 2: $N=25$, $n_{_2}=1$, $\beta_{_2}=0.75$, $\varepsilon_{_2}=0.01$, $\alpha=0.5$, $\kappa_{_{12}}=0.0$.\label{fig:3}}
\end{center}
\end{figure}

Fig.\ref{fig:3} shows the variation of $CV$ against the noise amplitude $\sigma_{_2}$ in the isolated layer 2 with a locally ($n_{_2}=1$) 
coupled ring network topology.
In the numerical computations, we choose $\varepsilon_{_2}=0.01$, a standard value used in the literature of the FHN neuron model.
For some values of the time delay $\tau_{_2}$ and the coupling force $\kappa_{_2}$, 
the round-bottom $CV$-curves show a non-monotonic dependence of $CV$ on $\sigma_{_2}$ 
with a well-pronounced deep minimum. The network thus exhibits a pronounced CR with coherent oscillations emerging 
in a very narrow interval of the noise amplitudes. 
For our model equation, the general behavior of the $CV$-curves in response to changes in the time-delayed coupling is the following: 
$CV$-curves are shifted to higher values with their minima occurring at higher noise amplitudes for stronger coupling forces and longer time delays.

In Fig.\ref{fig:3}\textbf{(a)}, for a fixed and weak coupling force of $\kappa_{_2}=0.1$, all the $CV$-curves obtained with different time delays 
($\tau_{_1}=0.0,\tau_{_2}=2.5, \tau_{_2}=5.0$, $\tau_{_2}=10.0$) show a deep minimum, indicating a high degree of coherence of the oscillation due to CR. 
We notice that for this weak coupling force, as $\tau_{_2}$ increases, 
the entire $CV$-curve is only slightly shifted up and to the right, thus slightly weakening CR and increasing the value of the noise 
amplitude at which maximal coherence is achieved.
For example, for $\tau_{_2}=0.0$, $CV_{min}=0.13$ at $\sigma_{_2}=0.82\times10^{-5}$ and for $\tau_{_2}=10.0$, $CV_{min}=0.20$ at $\sigma_{_2}=5.5\times10^{-5}$.

In Fig.\ref{fig:3}\textbf{(b)}, we now have a strong coupling force ($\kappa_{_2}=1.0$). The qualitative behavior due to longer time delays is the 
same as in Fig.\ref{fig:3}\textbf{(a)}, but is now much stronger, i.e., with strong coupling forces, 
the time delays have stronger effects on the $CV$-curves than with weak coupling forces. 
The $CV$-curves are shifted up to higher values with the minima occurring at much larger noise amplitudes than in  Fig.\ref{fig:3}\textbf{(a)}.
For example, for $\tau_{_2}=0.0$, $CV_{min}=0.14$ at $\sigma_{_2}=0.91\times10^{-5}$ and for $\tau_{_2}=7.0$, $CV_{min}=0.56$ at $\sigma_{_2}=4.6\times10^{-4}$.
We notice that for $\kappa_{_2}=1.0$ and $\tau_{_2}=7.0$, even though the red curve shows a non-monotonic dependence of $CV$
on $\sigma_{_2}$, it still has a relatively high minimum value of $CV_{min}=0.56$, indicating a poor CR. While for $\kappa_{_2}=0.1$ in 
Fig.\ref{fig:3}\textbf{(a)}, $\tau_{_2}=10.0$ still gives $CV_{min}=0.2$, indicating high coherence.

In Fig.\ref{fig:3}\textbf{(c)}, we now have a fixed and short time delay of $\tau_{_2}=0.1$ and we vary the coupling force $\kappa_{_2}$.
We observe that the coupling force between the neurons
does not have a significant effect on the coherence of the oscillations. 
The $CV$-curves at a short time delay of $\tau_{_2}=0.1$ show a well-pronounced deep minima with
$CV_{min}\approx0.13$ at $\sigma_{_2}\approx0.8\times10^{-5}$ for both weak and strong coupling forces. 
 
In Fig.\ref{fig:3}\textbf{(d)}, we fix the time delay at a relatively large value, $\tau_{_2}=7.0$, and we notice 
that the same values of the coupling force that could not change the values of $CV$ in Fig.\ref{fig:3}\textbf{(c)},
can now very strongly affect the $CV$-curves. For long time delays, stronger coupling forces between the neurons tend to 
decrease the coherence of the oscillations. In particular, for $\tau_{_2}=7.0$ and $\kappa_{_2}=1.5$, the pink $CV$-curve does not 
show a pronounced non-monotonic behavior and lies entirely above the line $CV=1.0$, indicating the presence of highly incoherent oscillations 
and the absence of CR.

Now, we turn our attention to the effects of different ring network topologies on CR. The numerical simulations carried out in Fig.\ref{fig:3} 
with a locally ($n_{_2}=1$) coupled ring network topology were correspondingly carried out with 
non-locally ($n_{_2}=3$ and $n_{_2}=6$) and  globally ($n_{_2}=12$) coupled 
ring network topologies, all with a network size of $N=25$. These simulations (not shown) depicted the same 
qualitative behavior as seen in Fig.\ref{fig:3}, i.e., CR deteriorating with stronger coupling forces and longer time delays. 
For a fixed coupling strength and time delay, we determined which of the ring network topologies (local, non-local, or global)
will deteriorate CR the most. 

\begin{figure}
\begin{center}
\includegraphics[width=4.5cm,height=4.25cm]{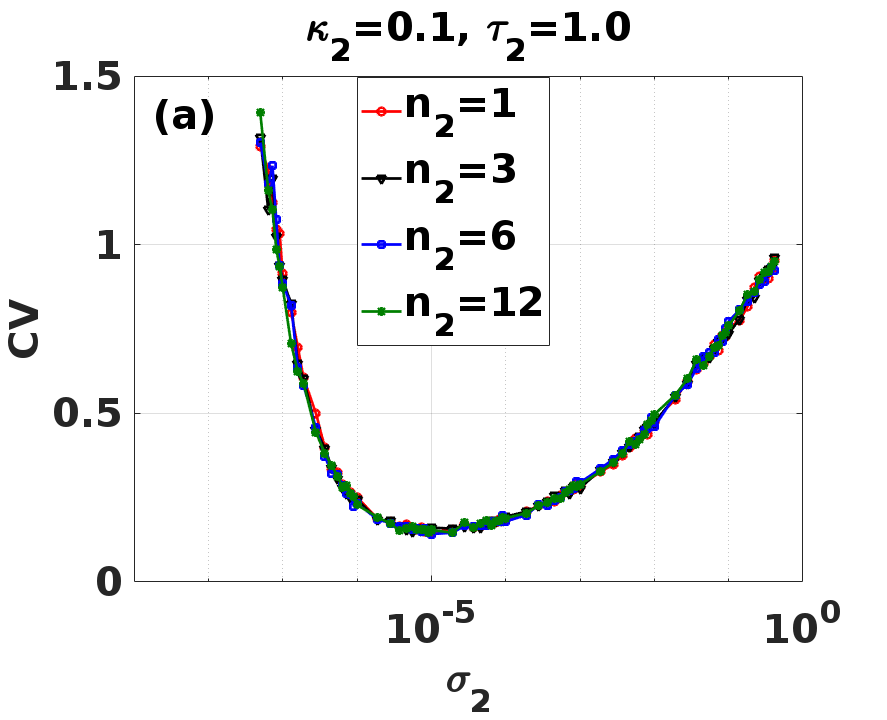}\includegraphics[width=4.5cm,height=4.25cm]{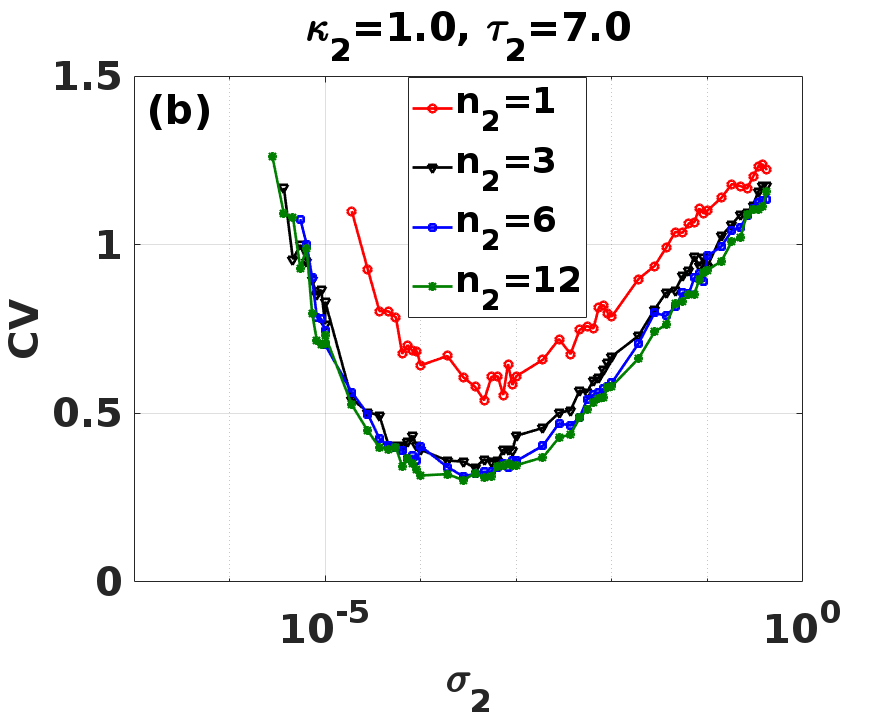}
\caption{(Color online) Coefficient of variation $CV$ against noise amplitude $\sigma_{_2}$ for different (local, nonlocal, global) ring network topologies in layer 2.
\textbf{(a):} changing the topology ($n_{_2}$) has no effect on the high coherence of oscillations in a weak coupling and short time delay regime.
\textbf{(b):} in a strong coupling and long time delay regime, the topology can significantly affect the coherence of the oscillations.
In this regime, the sparser the network is, the less coherent the oscillations are. 
Parameters of layer 2: $N=25$, $\beta_{_2}=0.75$, $\varepsilon_{_2}=0.01$, $\alpha=0.5$, $\kappa_{_{12}}=0.0$.\label{fig:4}}
\end{center}
\end{figure}

Fig.\ref{fig:4}\textbf{(a)} shows that in a regime of weak coupling force ($\kappa_{_2}=0.1$) and short time delay ($\tau_{_2}=1.0$),  locally, 
non-locally, and globally coupled ring network topologies allow a well-pronounced minimum in the value of $CV$.
In this regime, a change in the topology does not affect the $CV$ curve and leaves the high coherence of the oscillations unchanged for the network size
considered.
Here, for all the ring topologies: local ($n_{_2}=1$), 
non-local ($n_{_2}=3,6$), and global ($n_{_2}=12$), the corresponding $CV$-curves have the same 
minimum value of $CV_{min}\approx0.14$ at $\sigma_{_2}=0.91\times10^{-5}$. 
The reason behind this behavior lies in the fact that 
we have a \textit{weak} ($\kappa_{_2}\rightarrow0$) coupling limit between \textit{identical} oscillators, in which case there is very little interaction 
between oscillators (which behave in this limit as isolated oscillators). And because they are identical oscillators, 
they all show a very similar dynamic irrespective of the ring network topology.

The effects of the different ring network topologies on the $CV$-curve 
become significant only when we are in a regime of strong coupling forces ($\kappa_{_2}>0.5$) and long time delays $(\tau_{_2}>2.5)$. 
In Fig.\ref{fig:4}\textbf{(b)}, 
we fix the strong coupling force at $\kappa_{_2}=1.0$ and the long time delay at $\tau_{_2}=7.0$. We can see that all the $CV$-curves corresponding 
to different ring network topologies are shifted up to higher (compared to the corresponding $CV$-curves in Fig.\ref{fig:4}\textbf{(a)} 
with a weaker coupling and shorter delay) values as the ring network becomes sparser. 
Firstly, this confirms the fact that irrespective of the ring network 
topology (locally, non-locally or globally coupled), stronger coupling forces and longer time delays weaken CR which eventually 
disappears  at some sufficiently large values of these control parameters (see Fig.\ref{fig:3}\textbf{(d)}). 
Secondly, different ring network topologies can significantly affect the $CV$-curve 
only when the coupling forces are strong and the time delays are long. 

In Fig.\ref{fig:4}\textbf{(b)}, it can be clearly seen that  
the sparser the network is, the higher the $CV$-curve, and hence, the less pronounced is CR. 
This behavior can be explained by the fact that 
in a \textit{globally} coupled network of \textit{identical} oscillators 
with \textit{strong} coupling forces, the network relatively easily synchronizes, with the less coherent oscillators (having high $CV$ values)
synchronizing to adopt the dynamics of the more coherent ones (having low $CV$ values). 
This has an overall effect of lowering the averaged $CV$ of the entire network thus enhancing CR.
On the other hand, as the network becomes less dense (i.e., in a non-locally and especially locally coupled network), 
all the oscillators in the network can no longer so  easily synchronize (in particular, of course,  those which are not connected),
hence, the averaged $CV$ of the network is calculated with high $CV$ values (of less coherent oscillators which cannot easily synchronize) and 
low $CV$ values (of more coherent oscillators). This has the overall effect of shifting 
the averaged $CV$ to higher values and deteriorating CR as seen in Fig.\ref{fig:4}\textbf{(b)}. 

A poor or non-existent CR in layer 2 can therefore easily be obtained by imposing a strong coupling force with a long time delay 
in a locally coupled ring network topology. These are the three control parameter regimes ($\kappa_{_2}=1.0$ or $1.5$, $\tau_{_2}=7.0$, 
$n_{_2}=1$) in which we will set layer 2 before controlling (improve or induce) CR with multiplexing, CR and SISR in layer 1.

As we mentioned earlier, it has been shown in \cite{Semenova1 and Zakharova 2018} with another version of the FHN neuron model that
stronger coupling forces rather shift the $CV$-curve to lower values, thus enhancing CR. 
The version of the FHN neuron model used here is more general than 
that in \cite{Semenova1 and Zakharova 2018} (see also \cite{Maria et al 2017} which uses the same version as in \cite{Semenova1 and Zakharova 2018}) 
and shows a rather opposite effect, i.e., stronger coupling forces deteriorate CR. 
The reason behind this difference lies in the details of the equations of the two versions used.
For example, in the version used in \cite{Maria et al 2017}, the network of coupled neurons stays in the excitable regime when the 
coupling force is weak and time delay is short.
But as the strength of the coupling force and the length of the time delay increase, the network eventually switches from an excitable regime 
into a delayed-coupling-induced oscillatory regime via a saddle-node bifurcation. 
Therefore, increasing the coupling force tends to bring their version of the FHN model closer to the bifurcation threshold,
thus enhancing CR. In the current paper, only the slow variable equation of the neurons is different from that in
\cite{Maria et al 2017} and \cite{Semenova1 and Zakharova 2018}, but one can already see 
in Fig.\ref{fig:1a}\textbf{(a)} that increasing the strength of the coupling force and the length of the time delay within some interval
does not switch the network out of the excitable regime. As the coupling force becomes stronger and the time delay longer,
the network tends to move rather further away from the oscillatory regime, i.e., away from the 
Hopf bifurcation threshold. In Fig.\ref{fig:2}\textbf{(c)}, we even see 
that strong coupling and long time delays can switch the network from an oscillatory regime into a delayed-coupling-induced excitable regime;
this  effect is the exact opposite of that produced by the model used in \cite{Maria et al 2017} and \cite{Semenova1 and Zakharova 2018}.
In general, the further away we are from the (Hopf or saddle-node) bifurcation threshold, the more difficult it is for CR to occur. In such situations,
we need stronger noise amplitudes to have CR. This is why we see in Fig.\ref{fig:3} that as $\kappa_{_2}$ and $\tau_{_2}$ increase, the
$CV$-curves achieve their minima at larger values of $\sigma_{_2}$.

\subsubsection{\textbf{Self-Induced Stochastic Resonance}}
All previous works on SISR have considered single isolated oscillators.
The current work is thus the first to investigate SISR in a network of coupled oscillators. 
Here, we investigate the effects of time delay ($\tau_{_1}$), coupling strength ($\kappa_{_1}$), 
and different ring network topologies ($n_{_1}$) on SISR in a layer network of coupled FHN oscillators.
We compare the behavior of SISR to that of CR when these parameters are changed.
Understanding the effects of $\tau_{_1}$, $\kappa_{_1}$, and $n_{_1}$ on SISR in layer 1 will guide us in controlling CR in layer 2 with
SISR via the multiplexing of both layers.

\begin{figure}
\begin{center}
\includegraphics[width=4.5cm,height=4.5cm]{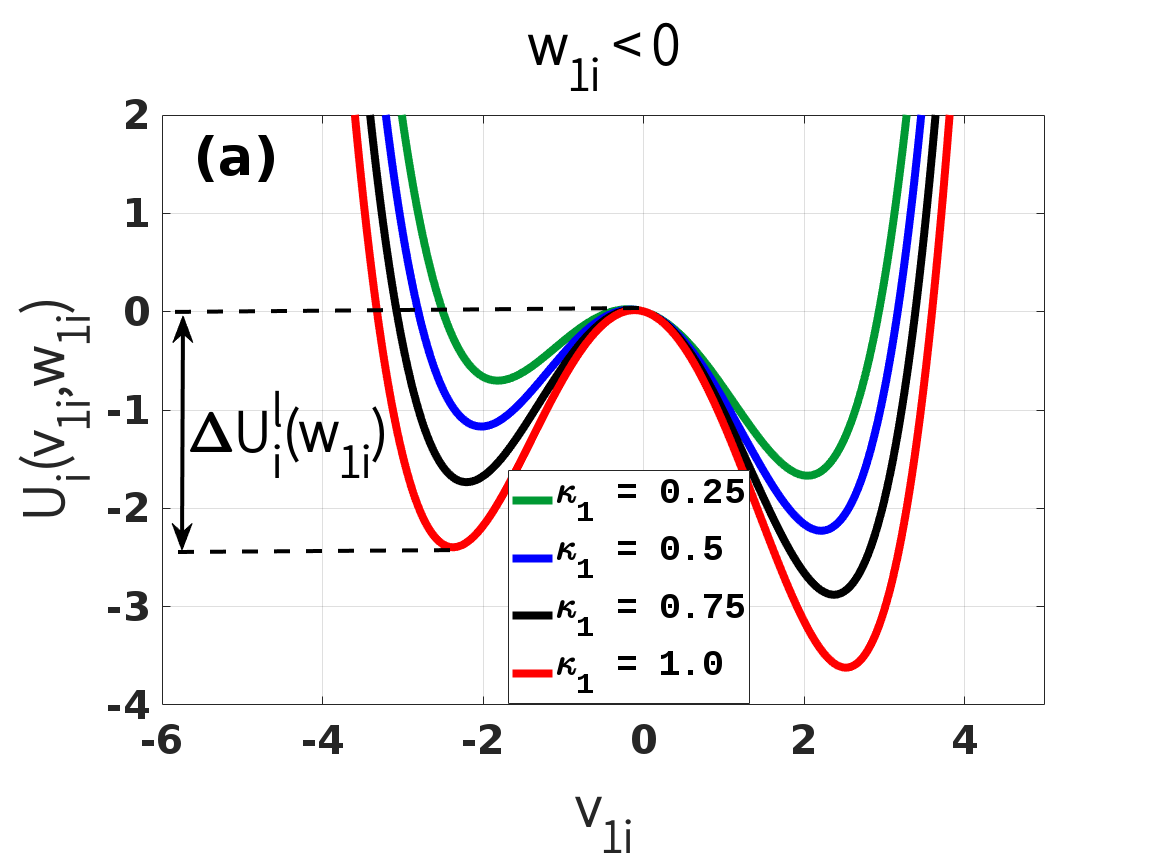}\includegraphics[width=4.5cm,height=4.5cm]{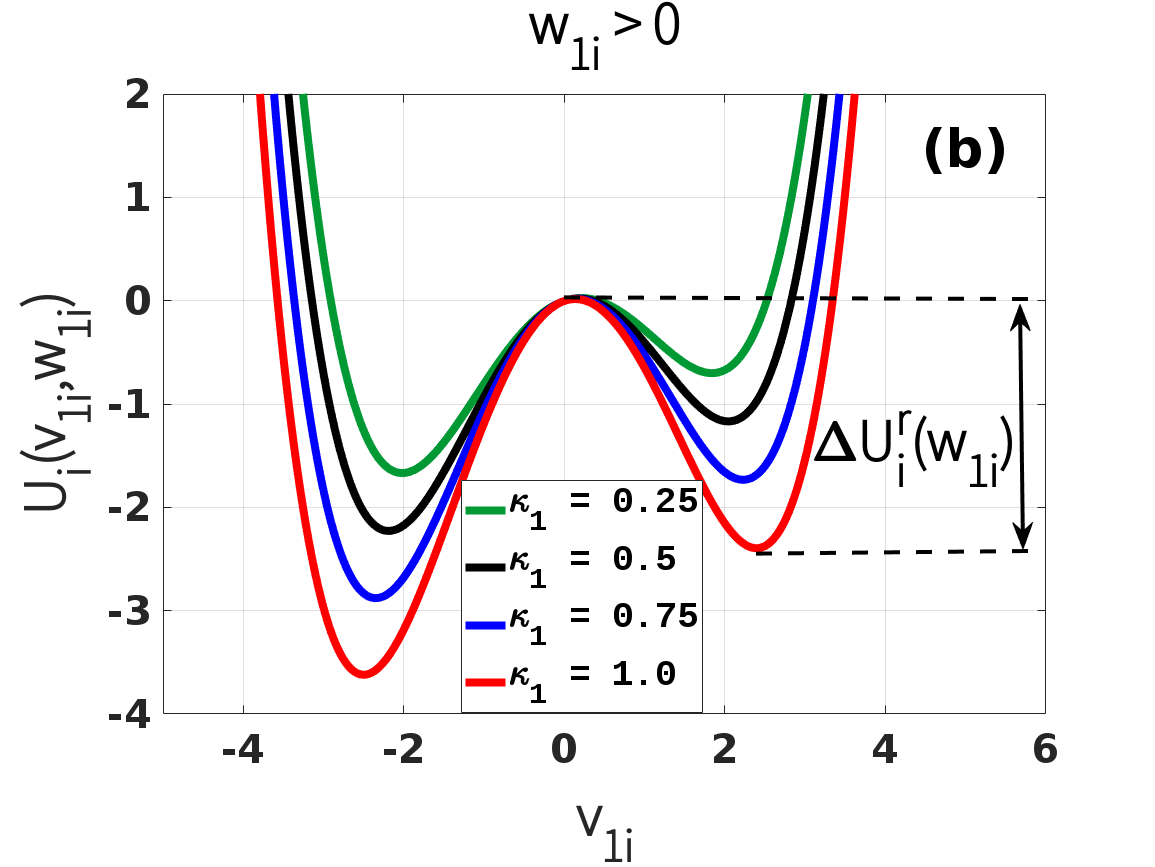}
\includegraphics[width=4.5cm,height=4.5cm]{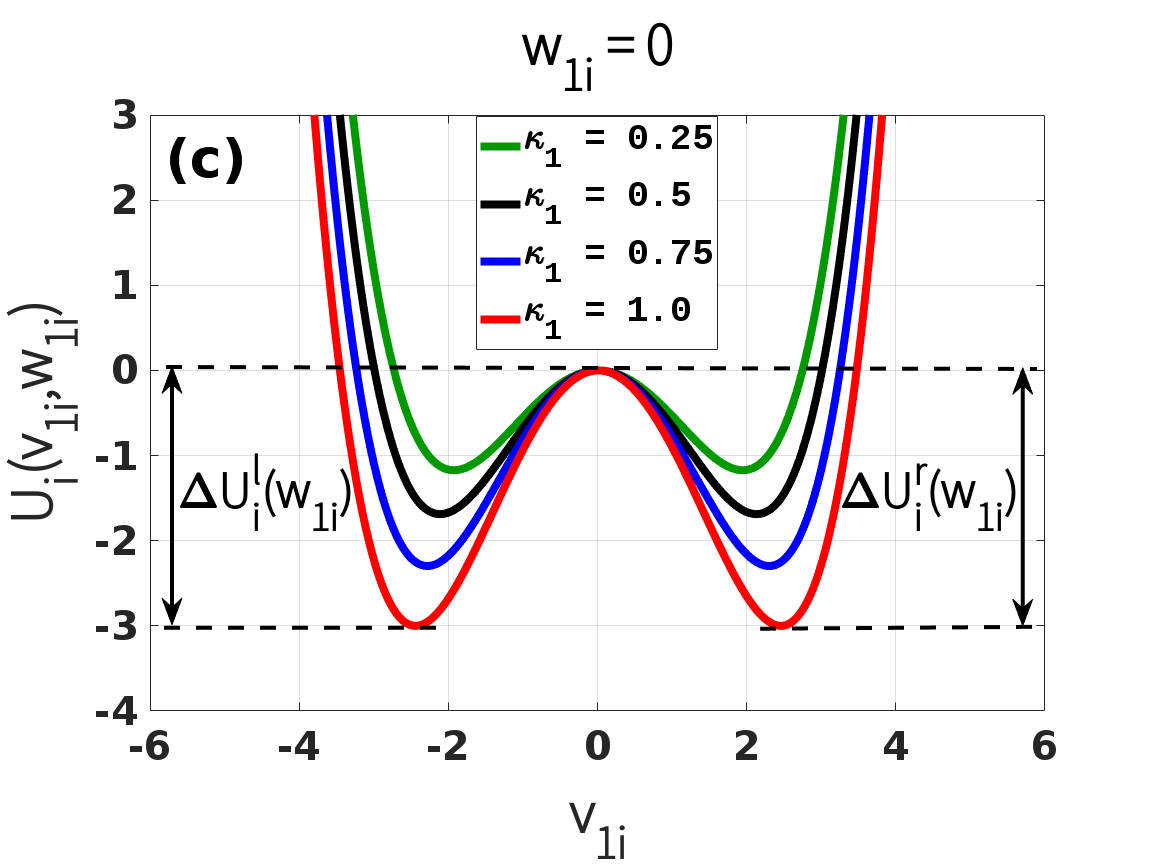}
\caption{(Color online) Landscapes of the interaction potential $U_i(v_{_{1i}},w_{_{1i}})$ in Eq.\ref{eq:5b} for a locally ($n_{_1}=1$) coupled ring network topology
with the energy barriers indicated in the asymmetric ($w_{_{1i}}\neq0$) cases \textbf{(a)} and $\textbf{(b)}$ and
symmetric case \textbf{(c)}. The stronger the intra-layer coupling force $\kappa_{_1}$ is, the deeper the 
energy barrier functions $\triangle U_i^l(w_{_{1i}})$ and 
$\triangle U_i^r(w_{_{1i}})$ are. In particular, when $w_{_{1i}}=0$, $\triangle U_i^l(w_{_{1i}})=\triangle U_i^r(w_{_{1i}})$, 
and both energy barriers achieve the deepest height at the strongest coupling force $\kappa_{_1}$.
The saddle point and the left and right minima of the interaction potential are located at
$v_{_{1i}}=v^*_m(w_{_{1i}})$, $v_{_{1i}}=v^*_l(w_{_{1i}})$, and $v_{_{1i}}=v^*_r(w_{_{1i}})$, respectively, i.e., $v^*_l(w_{_{1i}})<v^*_m(w_{_{1i}})<v^*_r(w_{_{1i}})$
\label{fig:5a}}
\end{center}
\end{figure}

In the isolated layer 1 with $\sigma_{_3}=0$, and in the limit as $\varepsilon_{_1}\rightarrow0$, 
each neuron equation in this layer reduces to coupled Langevin equations of the form:
\begin{eqnarray}\label{eq:5b}
\begin{split}
\left\{\begin{array}{lcl}
\displaystyle{ dv_{_{1i}}}&=&\displaystyle{ -\frac{\partial U_i(v_{_{1i}},w_{_{1i}})}{\partial v_{_{1i}}}dt+\sigma_{_1} dW_{_{1i}}},\\[5.0mm]
U_i(v_{_{1i}},w_{_{1i}})&=&\frac{1}{12}v_{_{1i}}^4-\frac{1}{2}v_{_{1i}}^2+v_{_{1i}}w_{_{1i}}\\[3.0mm]
&-&\frac{\kappa_{_1}}{2n_{_1}}\sum\limits_{j=i-n_{_1}}^{i+n_{_1}}\Big(v_{_{1i}}(t)v_{_{1j}}(t-\tau_{_1})-\frac{1}{2}v_{_{1i}}(t)^2\Big),
\end{array}\right.
\end{split}
\end{eqnarray}
where the interaction potentials $U_i(v_{_{1i}},w_{_{1i}})$ ($i=1,...,N$) viewed as a function of $v_{_{1i}}$ with $w_{_{1i}}$ nearly constant, are 
double-well potentials. Fig.\ref{fig:5a} shows the landscape of these interaction potentials. 

When $w_{_{1i}}<0$, 
$U_i(v_{_{1i}},w_{_{1i}})$ is asymmetric with the shallower well at the left. In this case, the neuron is close to the homogeneous stable  
fixed point at $(v^*_{_{1i}},w^*_{_{1i}})=(-1.003975,-0.666651)$ and a spike consists of jumping over the left energy barrier 
$\triangle U_i^l(w_{_{1i}})$ into the right well, see Fig.\ref{fig:5a}\textbf{(a)}. 

When $w_{_{1i}}>0$, $U_i(v_{_{1i}},w_{_{1i}})$ is also asymmetric. In this case,
the neuron has spiked and a return to the quiescent state (the homogeneous
stable fixed point) consists of jumping over the right energy barrier $\triangle U_i^r(w_{_{1i}})$ into the left well, see Fig.\ref{fig:5a}\textbf{(b)}.
When $w_{_{1i}}=0$, $U_i(v_{_{1i}},w_{_{1i}})$ is symmetric with $\triangle U_i^l(w_{_{1i}})=\triangle U_i^r(w_{_{1i}})$, in which case the neuron is half 
way between the quiescent state and the spike state, see Fig.\ref{fig:5a}\textbf{(c)}. 

We can also see in Fig.\ref{fig:5a} that the intra-layer coupling force $\kappa_{_1}$ does not change the symmetry (or asymmetry) of 
the interaction potential. It only changes the depth of the energy barriers. 
The stronger $\kappa_{_1}$ is, the deeper the energy barrier functions $\triangle U_i^l(w_{_{1i}})$ and 
$\triangle U_i^r(w_{_{1i}})$ defined in Eq.\ref{eq:5a} are. 
Later, this particular behavior shall explain why SISR is deteriorated by stronger intra-layer coupling forces.
We choose the parameters of the coupled neurons in Eq.\ref{eq:5b} such that they satisfy the conditions necessary for the occurrence of SISR. 
These conditions which are adapted from those of a single isolated FHN neuron 
\cite{Lee DeVille et al. 2005,Yamakou and Jost 2018} to include the time-delayed 
coupling between the neurons coupled in a ring network topology are such that:
\begin{eqnarray}\label{eq:4}
\begin{split}
\left\{\begin{array}{lcl}
\displaystyle{ \lim \limits_{(\varepsilon_{_1},\sigma_{_1})\rightarrow (0,0)}\frac{\sigma_{_1}^2}{2}\log_e(\varepsilon_{_1}^{-1})\in 
\Big(\triangle U_i^l(w^*_{_{1i}}),F(\kappa_{_1},\tau_{_1},n_{_1})\Big)},\\[4.0mm]
\displaystyle{ \lim \limits_{(\varepsilon_{_1},\sigma_{_1})\rightarrow (0,0)}\frac{\sigma_{_1}^2}{2}\log_e(\varepsilon^{-1}_{_1})=\mathcal{O}(1)},\\[5.0mm]
\beta_{_1}-\beta_{_h}(\varepsilon_{_1})>0,\\[2.0mm]
\sigma_{_3}=0,
\end{array}\right.
\end{split}
\end{eqnarray}
where
\begin{eqnarray}\label{eq:5a}
\begin{split}
\left\{\begin{array}{lcl}
F(\kappa_{_1},\tau_{_1},n_{_1}):=\Big\{(\kappa_{_1},\tau_{_1},n_{_1}):\triangle U_i^l(w_{_{1i}})=\triangle U_i^r(w_{_{1i}})\Big\},\\[3.0mm]
\triangle U_i^l(w_{_{1i}}):=U_i\big(v^*_m(w_{_{1i}}),w_{_{1i}}\big)- U_i\big(v^*_l(w_{_{1i}}),w_{_{1i}}\big),\\[2.0mm]
\triangle U_i^r(w_{_{1i}}):=U_i\big(v^*_m(w_{_{1i}}),w_{_{1i}}\big)- U_i\big(v^*_r(w_{_{1i}}),w_{_{1i}}\big),
\end{array}\right.
\end{split}
\end{eqnarray}
with
\begin{equation}\label{eq:5c}
\begin{split}
v^*_{l,m,r}(w_{_{1i}}):=\Big\{v_{_{1i}}: v_{_{1i}}-\displaystyle{\frac{v_{_{1i}}^3}{3}}-w_{_{1i}}\\
+\frac{\kappa_{_1}}{2n_{_1}}\sum\limits_{j=i-n_{_1}}^{i+n_{_1}}\Big(v_{_{1j}}(t-\tau_{_1})-v_{_{1i}}(t)\Big)=0,\\
v^*_l(w_{_{1i}})<v^*_m(w_{_{1i}})<v^*_r(w_{_{1i}})\Big\}.
\end{split}
\end{equation}
The energy barrier functions $\triangle U_i^{l,r}(w_{_{1i}})$ are obtained from the 
potential $U_i(v_{_{1i}},w_{_{1i}})$  by taking the difference between the potential function value
at the saddle point $v^*_m(w_{_{1i}})$ and at the local minima $v^*_{l,r}(w_{_{1i}})$ of the double potential $U_i(v_{_{1i}},w_{_{1i}})$, 
respectively \cite{Yamakou and Jost 2018}. 
$\triangle U_i^l\big(w^*_{_{1i}}\big)$  (which has to be crossed to induce a spike) 
is the value of the left energy barrier function at the $w_{_{1i}}$-coordinate of the 
homogeneous stable fixed point $\big(v^*_{_{1i}}(\kappa_{_1},\tau_{_1},n_{_1}),w^*_{_{1i}}(\kappa_{_1},\tau_{_1},n_{_1})\big)$. 
This is where $U_i^l\big(w^*_{_{1i}}(\kappa_{_1},\tau_{_1},n_{_1})\big)$ gets its $\kappa_{_1}$, $\tau_{_1}$, and $n_{_1}$ dependence. 
We note that $n_{_1}$ can only change the value of this fixed point because $v^*_{_{1i}}$ and $w^*_{_{1i}}$ depend on $n_{_1}$. 
But changing $n_{_1}$ (or even the network size $N$) does not take the network out of the excitable regime
as long as $\beta_{_1}-\beta_{_h}(\varepsilon_{_1})>0$, see Fig.\ref{fig:2}\textbf{(b)}.

\begin{figure}
\begin{center}
\includegraphics[width=4.5cm,height=4.25cm]{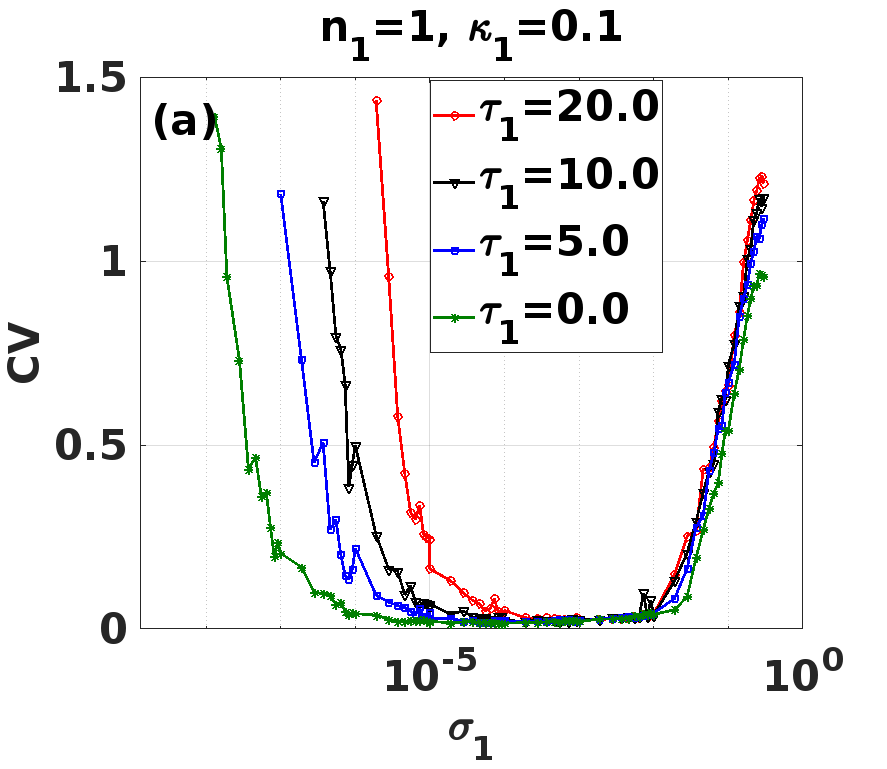}\includegraphics[width=4.5cm,height=4.25cm]{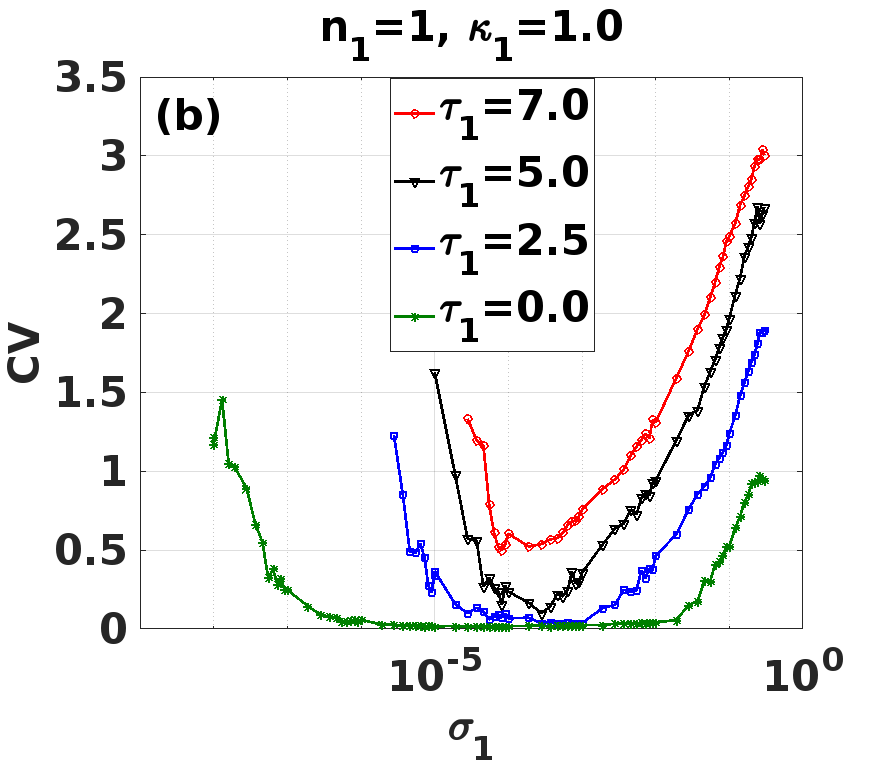}
\includegraphics[width=4.5cm,height=4.25cm]{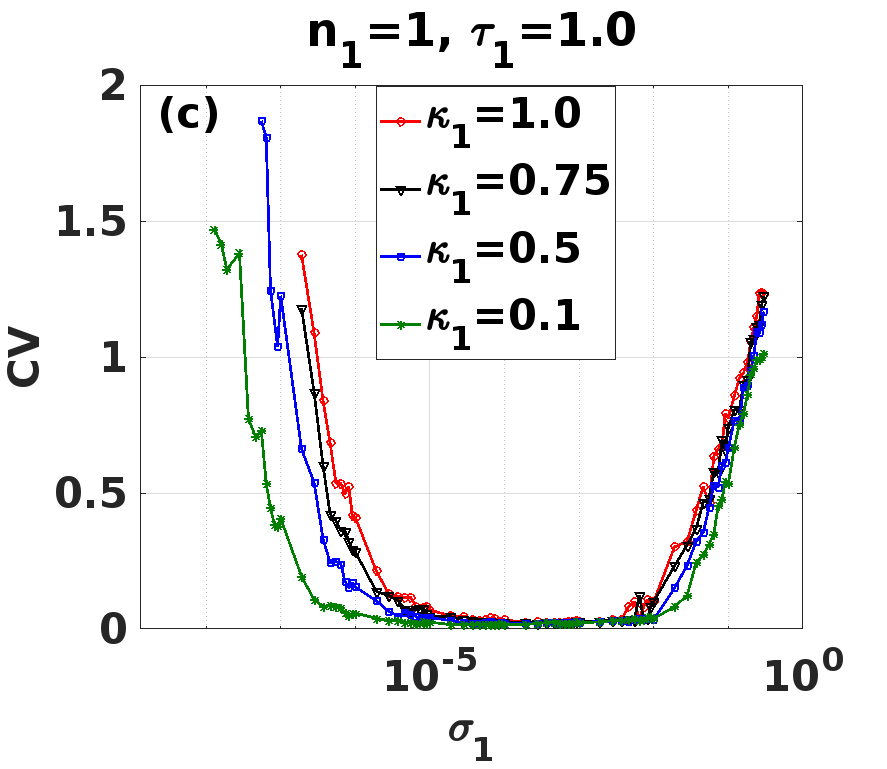}\includegraphics[width=4.5cm,height=4.25cm]{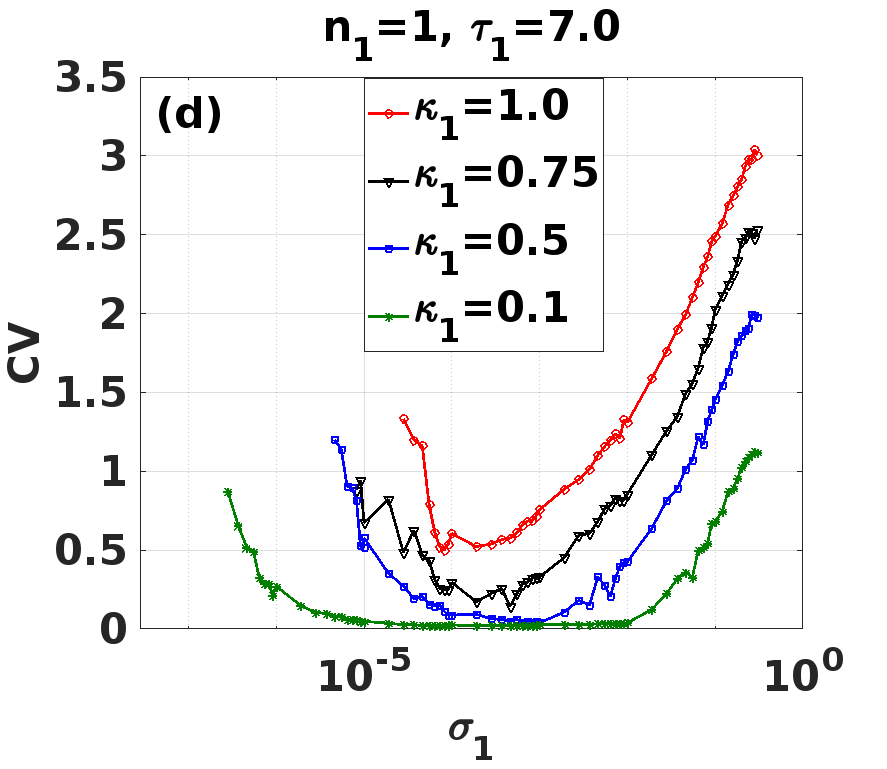}
\caption{(Color online) Coefficient of variation $CV$ against noise amplitude $\sigma_{_1}$ of layer 1 in isolation.
Like CR in Fig.\ref{fig:3}, SISR is weakened by stronger coupling $\kappa_{_1}$ and longer delays $\tau_{_1}$.
SISR is, however, much less sensitive to these parameters than CR. The flat-bottom of the $CV$-curves indicates 
that a wider range of noise amplitude can induce coherent oscillations during SISR, unlike CR in Fig.\ref{fig:3} with round-bottom $CV$-curves. 
For the same time-delayed couplings, oscillation due to SISR are more coherent (lower $CV$-curves) than those due to CR. 
In \textbf{(b)} and \textbf{(d)} with a strong coupling force and long time delay regime, 
the maximum noise amplitude below which SISR occurs is controllable, but fixed in \textbf{(a)} and \textbf{(c)} with a weak coupling force
and short time delay regime.
Parameters of layer 1:  $N=25$, $n_{_1}=1$, $\beta_{_1}=0.75$, $\varepsilon_{_1}=0.0005$, $\alpha=0.5$, $\kappa_{_{12}}=0.0.$ \label{fig:5}}
\end{center}
\end{figure}

Fig.\ref{fig:5} shows the variation of $CV$ against the noise amplitude $\sigma_{_1}$ in layer 1 with a locally ($n_{_1}=1$) coupled ring network topology.
In the numerical computations, we choose $\varepsilon_{_1}=0.0005$ because unlike CR (with $\varepsilon_{_2}=0.01$), SISR can only occur 
in the singular limit $\varepsilon_{_1}\rightarrow0$ imposed in Eq.\eqref{eq:4}. 
Furthermore, we note that the noise amplitude $\sigma_{_3}$ in 
Eq.\eqref{eq:1} which is attached to the slow variable ($w_{_{1i}}$) equations is set to zero in the conditions necessary for SISR in Eq.\eqref{eq:4}.
We recall that for the FHN model, SISR requires the noise term to be attached to the fast variable ($v_{_{1i}}$) equation, while
CR requires the noise term to be attached to the slow variable ($w_{_{1i}}$) equation \cite{Lee DeVille et al. 2005}.
Therefore, in our study, the noise amplitude $\sigma_{_3}$ will only be used in the multiplex network setting,
when we want to investigate  the enhancement of CR in layer 2 by the occurrence of CR in layer 1.
This will later allow us to compare the enhancement capabilities of CR and SISR.

In Fig.\ref{fig:5}, the general behavior of SISR is the following: In the regime of weak coupling forces $\kappa_{_1}$ and short time delays $\tau_{_1}$,
the $CV$-curves remain very low, indicating a high degree of coherence of the oscillations. 
This high degree of coherence is gradually lost as one moves to stronger coupling forces and longer time delay. 
This general behavior has also been observed with CR in Fig.\ref{fig:3}. 
The reason for this behavior for the case SISR is the same as for CR, i.e., where larger values 
of $\kappa_{_1}$ and $\tau_{_1}$ bring the system further away from the Hopf bifurcation threshold,
making the noise-induced oscillations to be less coherent, as a larger noise amplitude is now required to induce oscillations. 
Moreover, the $CV$-curves of SISR in Fig.\ref{fig:5} show oscillations that are much more coherent for 
a wider range of noise amplitude $\sigma_{_1}$ than the $CV$-curves of CR in Fig.\ref{fig:3}, which are rather relatively higher with the minimum 
occurring at single value of the noise amplitude $\sigma_{_2}$. This confirms the more robust nature of SISR compared to CR.

The response of the SISR to changes in the coupling force and time delay in Fig.\ref{fig:5} can also be explained in terms of the 
interaction potential $U_i(v_{_{1i}},w_{_{1i}})$ in Fig.\ref{fig:5a}. We observe in Fig.\ref{fig:5a} that for a fixed 
ring network topology and time delay, as the coupling force $\kappa_{_1}$ increases, 
the energy barriers $\triangle U_i^l(w_{_{1i}})$ and $\triangle U_i^r(w_{_{1i}})$ become deeper. In particular, when $w_{_{1i}}<0$, 
the trajectory is in the left potential well and as $\kappa_{_1}$ becomes stronger ($0.25, 0.5, 0.75, 1.0$), the left energy barrier 
$\triangle U_i^l(w_{_{1i}})$ becomes deeper (hence the trajectory at the bottom of the well get closer to the homogeneous  stable  
fixed point at $w^*_{_{1i}}=-0.666651$). Thus, the deeper the left energy barrier $\triangle U_i^l(w_{_{1i}})$ 
is (in other words, the stronger the coupling force $\kappa_{_1}$ is), the closer is the trajectory to the stable fixed point and the further away is 
the system from the oscillatory regime. 

For the trajectory to jump over this deep left energy barrier, a stronger noise amplitude is 
thus needed. This is why in Fig.\ref{fig:5} as $\kappa_{_1}$ increases, the left branch of the $CV$-curve is shifted to the right, 
meaning that stronger noise amplitudes are required to induced frequent spiking (i.e., frequent escape from the deep left energy barrier).
Furthermore, we can see (from Fig.\ref{fig:5}\textbf{(d)}) that at longer time delay $\tau_{_1}$, this effect
(the shifting of the left branch of the $CV$-curve to the right) is more pronounced than 
in Fig.\ref{fig:5}\textbf{(c)} with a shorter time delay. This is because in Eq.\ref{eq:5b}, the longer the time delay is ($\tau_{_1}\gg0$), 
the further away is the quantity $\big[v_{_{1i}}(t)v_{_{1j}}(t-\tau_{_1})-v_{_{1i}}(t)^2\big]$ from zero (given that the oscillators are identical),
and hence, the stronger the effect of the coupling force
$\kappa_{_1}$ on the interaction potential, the energy barriers functions and consequently, on the $CV$-curves. 
Otherwise, if $\tau_{_1}\rightarrow0$, then because the oscillators are identical,
$\big[v_{_{1i}}(t)v_{_{1j}}(t-\tau_{_1})-v_{_{1i}}(t)^2\big]\rightarrow0$, and $\kappa_{_1}$ will have little effect on the interaction potential, the energy barriers, 
and on the $CV$-curves. This is why the coupling force $\kappa_{_1}$  
has a stronger effect on SISR only when $\tau_{_1}$ gets longer, and vice versa.

In Fig.\ref{fig:5}\textbf{(a)}, for a fixed and weak coupling force $\kappa_{_1}=0.1$, all the flat-bottom $CV$-curves obtained with different time delays 
($\tau_{_1}=0.0,\tau_{_1}=5.0, \tau_{_1}=10.0$, $\tau_{_1}=20.0$) show a deep and broad minimum, indicating a high degree of coherence for a wide range of the 
noise amplitude. We notice that for this weak coupling force, as $\tau_{_1}$ increases, 
the $CV$-curve is not shifted to higher values as in the case of CR in Fig.\ref{fig:3}\textbf{(a)}.
Here, only the left branch of the $CV$-curve is significantly shifted to the right.
This means that in the weak coupling regime, the coherence of the oscillations due to SISR is not affected as the delay becomes longer, but 
the coherence is achieved only at larger noise amplitudes $\sigma_{_1}$. 
In Fig.\ref{fig:5}\textbf{(a)}, we have the same minimum
value of $CV_{min}\approx0.012$ for: $\tau_{_1}=0.0$ with $\sigma_{_1}\in\big(2.7\times10^{-7},2.8\times10^{-2}\big)$; 
$\tau_{_1}=5.0$ with $\sigma_{_1}\in\big(9.9\times10^{-7},1.9\times10^{-2}\big)$;
$\tau_{_1}=10.0$ with $\sigma_{_1}\in\big(6.3\times10^{-6},1.85\times10^{-2}\big)$; 
and $\tau_{_1}=20.0$ with $\sigma_{_1}\in\big(2.8\times10^{-5},1.8\times10^{-2}\big)$. Notice that the lower bound of the 
intervals increases as the delay increases while the upper bounds are almost fixed.

In Fig.\ref{fig:5}\textbf{(b)}, we switch to a strong coupling ($\kappa_{_1}=1.0$) regime. We see that in contrast to the weak coupling regime, 
increasing the length of the time delay squeezes the left and right branches of the $CV$-curves into a smaller noise interval, while shifting
the curves to higher values, thus decreasing the coherence of the oscillation. Here, we have different noise intervals for different minima of $CV$.
We have $CV_{min}=0.014$ at $\tau_{_1}=0.0$ for $\sigma_{_1}\in\big(2.7\times10^{-7},2.8\times10^{-2}\big)$;  
$CV_{min}=0.071$ at $\tau_{_1}=2.5$ for $\sigma_{_1}\in\big(2.8\times10^{-5},1.0\times10^{-3}\big)$; and in the last two cases, the noise intervals in which we have the highest coherence
have contracted to points with $CV_{min}=0.16$ at $\sigma_{_1}=2.8\times10^{-4}$ for $\tau_{_1}=5.0$; 
and $CV_{min}=0.51$ at $\sigma_{_1}=8.2\times10^{-5}$ for  $\tau_{_1}=7.0$. 

In Fig.\ref{fig:5}\textbf{(b)}, firstly, we notice that in the strong coupling regime, the effect of the time delay on the coherence 
of the oscillations becomes significant unlike
when the coupling is weak as in Fig.\ref{fig:5}\textbf{(a)}. In Fig.\ref{fig:5}\textbf{(b)}, at only $\tau_{_1}=7.0$, $CV_{min}$ is already
above 0.5, whereas in the weak coupling regime in Fig.\ref{fig:5}\textbf{(a)}, even at $\tau_{_1}=20.0$, we have $CV_{min}\approx0.012$.

Secondly, in the weak coupling and short time delay regimes
(i.e., in Fig.\ref{fig:5}\textbf{(a)} and \textbf{(c)}, respectively), the upper bound of the noise interval for which the 
$CV$-curves achieve their minima is almost fixed. Here, only the lower bound of the noise intervals is shifted to the right. 
Whereas in the strong coupling and long time delay regimes (i.e., in Fig.\ref{fig:5}\textbf{(b)} and \textbf{(d)}, respectively), 
both the lower and upper bounds of the noise intervals are shifted respectively to the right and to the left 
as the $\tau_{_1}$ and $\kappa_{_1}$ increase. This has the overall effect of shrinking the noise interval in which the $CV$-curves achieve their minima
to a single value of $\sigma_{_1}$.

\begin{figure}
\begin{center}
\includegraphics[width=4.5cm,height=4.25cm]{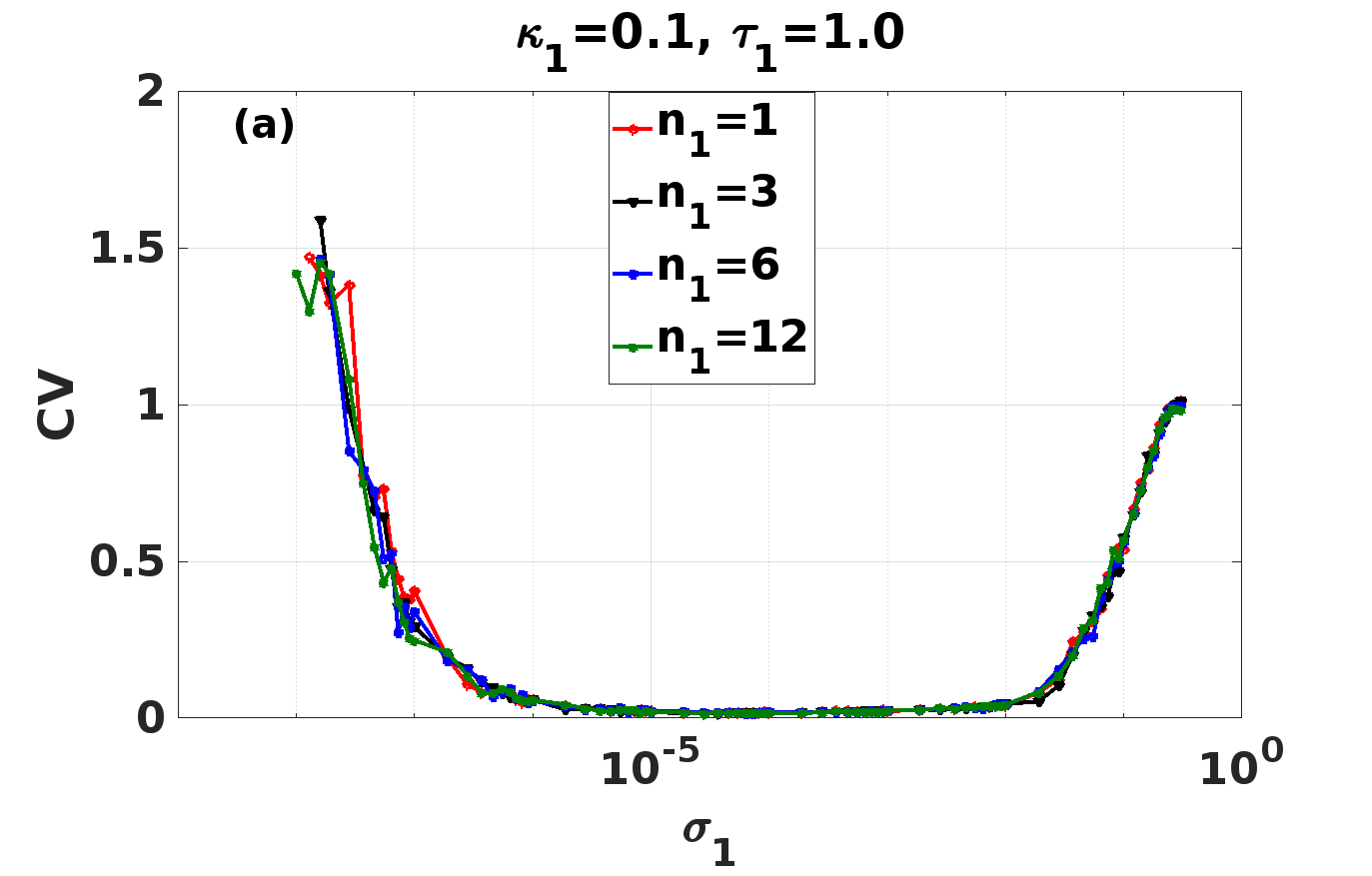}\includegraphics[width=4.5cm,height=4.25cm]{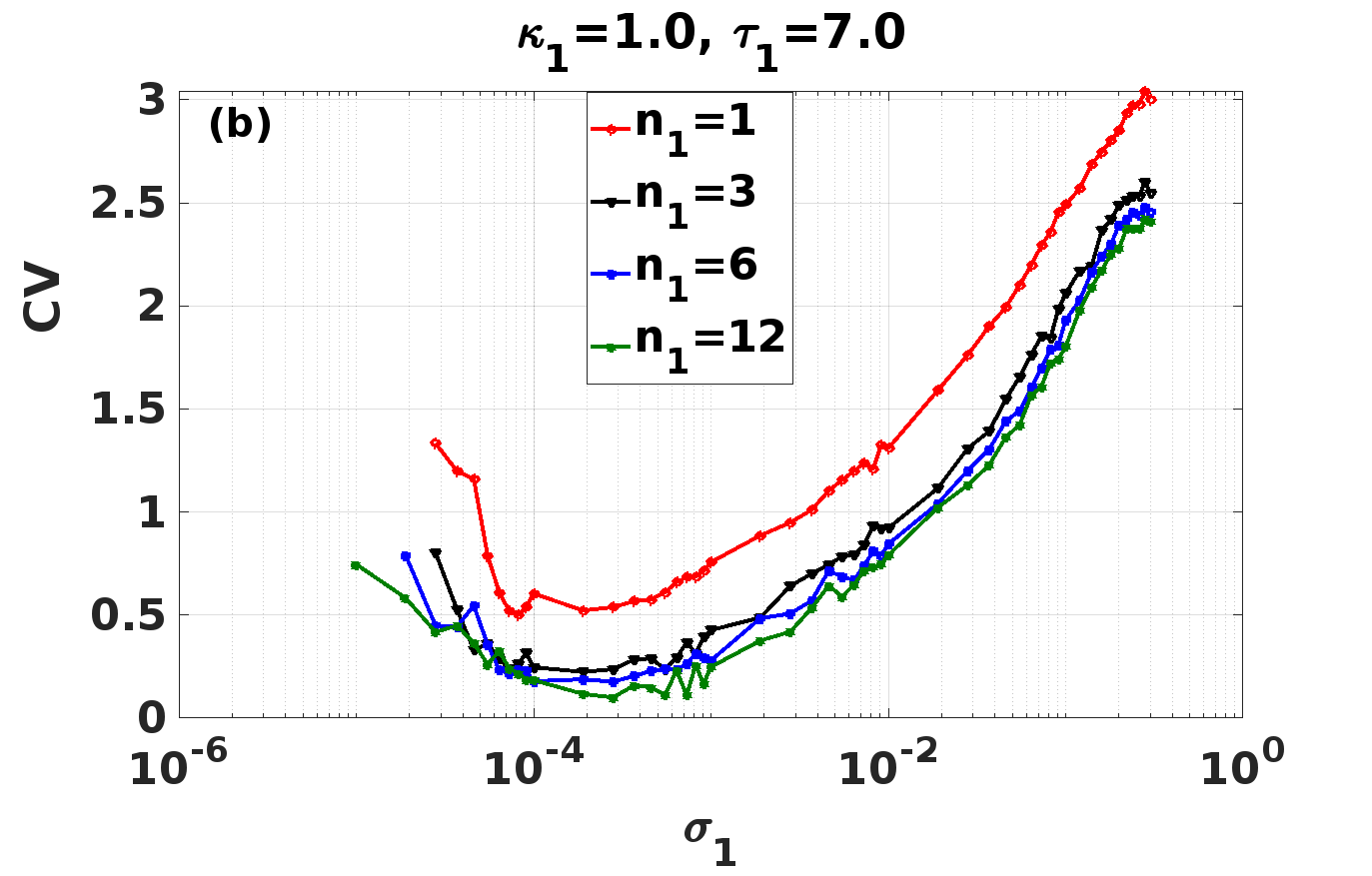}
\caption{(Color online) Coefficient of variation $CV$ against noise amplitude $\sigma_{_1}$ for different (local, nonlocal, global) ring network topologies.
\textbf{(a):} changing the topology ($n_{_1}$) has no effect on the high coherence of oscillations in a weak coupling and short time delay regime.
\textbf{(b):} in a strong coupling and long time delay regime, the ring topology can significantly affect the coherence of the oscillations.
In this regime, the sparser the network is, the less coherent the oscillations due to SISR are. 
Parameters of layer 1: $N=25$, $\beta_{_1}=0.75$, $\varepsilon_{_1}=0.0005$, $\alpha=0.5$, $\kappa_{_{12}}=0.0$.\label{fig:6}}
\end{center}
\end{figure}

To explain this behavior, we obtain from the first condition of Eq.\ref{eq:4} the minimum and maximum noise amplitudes between which SISR occurs as: 
\begin{eqnarray}\label{eq:7}
\begin{split}
\left\{\begin{array}{lcl}
\displaystyle{\sigma^{min}_1=\sqrt{\frac{2\triangle U_i^l\big(w^*_{_{1i}}(\kappa_{_1},\tau_{_1},n_{_1})\big)}{\log_e(\varepsilon_{_1}^{-1})}}},\\[5.0mm]
\displaystyle{\sigma^{max}_1=\sqrt{\frac{2F\big(\kappa_{_1},\tau_{_1},n_{_1}\big)}{\log_e(\varepsilon_{_1}^{-1})}}}.
\end{array}\right.
\end{split}
\end{eqnarray}
We observe that $\sigma^{min}_1$ depends on the fixed point coordinate $w^*_{_{1i}}(\kappa_{_1},\tau_{_1},n_{_1})$ which in turn also depends on $\kappa_{_1},\tau_{_1},$ and $n_{_1}$. 
Therefore, changing $\kappa_{_1},\tau_{_1},$ and $n_{_1}$ will change the value of $w^*_{_{1i}}(\kappa_{_1},\tau_{_1},n_{_1})$ which will in turn change the value of $\sigma^{min}_1$ 
via the energy barrier function $\triangle U_i^l(w^*_{_{1i}})$.
Numerical computations show that $\sigma^{min}_1$ increases as $\kappa_{_1}$ and $\tau_{_1}$ increase (see Fig.\ref{fig:5}) and as $n_{_1}$ decreases (see Fig.\ref{fig:6}\textbf{(b)}).

On the other boundary, $\sigma^{max}_1$ does not depend on the coordinates of the homogeneous stable fixed point, but on the complicated function 
$F(\kappa_{_1},\tau_{_1},n_{_1})$, completely defined by Eq.\ref{eq:5a} and Eq.\ref{eq:5c}. In Fig.\ref{fig:5}\textbf{(a)} and \textbf{(c)} (i.e., 
in the weak coupling and short time delay regimes, respectively), we notice that $\sigma^{max}_1=\mathcal{O}(10^{-2})$ is almost fixed for all 
the values of the time delay and coupling strength used. This fixation of the upper bound of the noise interval in which SISR occurs 
was already observed in a single isolated FHN neuron \cite{Yamakou and Jost 2018}. In the case of a single isolated FHN neuron, the  function in 
Eq.\ref{eq:5a} does not depend on any system parameter and takes a simple constant value $F=\frac{3}{4}$. 
This implies (for a fixed $\varepsilon_{_1}=0.0005$) a fixed value for $\sigma^{max}_1=\big(3/2\cdot\log_e(\varepsilon_{_1}^{-1})\big)^{1/2}$. 

In the current case of coupled FHN neurons, the fixation of the upper bound of the noise interval in which SISR occurs can therefore only be observed
if $F(\kappa_{_1},\tau_{_1},n_{_1})\rightarrow C$, where $C$ is a constant. In particular,  in a weak coupling ($\kappa_{_1}\rightarrow0$) regime and 
short time delay ($\tau_{_1}\rightarrow0$) regime (or more precisely, $v_{_{1j}}(t-\tau_{_1})-v_{_{1i}}(t)\rightarrow0$ as $\tau_{_1}\rightarrow0$, 
because all the oscillators are identical), $F(\kappa_{_1},\tau_{_1},n_{_1})\rightarrow\frac{3}{4}$. 
In these regimes (see Fig.\ref{fig:5}\textbf{(a)} and \textbf{(c)}), we observe that $\sigma_{_1}^{max}\approx10^{-2}$, that is the value
obtained in \cite{Yamakou and Jost 2018} for the case of a single isolated ($\kappa_{_1}=0$) FHN neuron.

In the strong ($\kappa_{_1}\gg0$) coupling and long ($\tau_{_1}\gg0\Rightarrow v_{_{1j}}(t-\tau_{_1})-v_{_{1i}}(t)\neq0$) time delay regimes
(Fig.\ref{fig:5}\textbf{(b)} and \textbf{(d)}), the function $F$ in Eq.\ref{eq:5a} is now strongly 
modified by $n_{_1}$ and the large values of $\kappa_{_1}$ and $\tau_{_1}$.
This is why in these regimes, the upper bound $\sigma_{_1}^{max}$ of the noise interval in which SISR occurs is not longer fixed, but shifted to the left as
$\kappa_{_1}$ and $\tau_{_1}$ take on larger values.
This left-shifting of the upper bound of the noise interval cannot occur 
in an isolated FHN neuron because it owes its occurrence to the (strong) coupling and (long) time delays in a network of coupled FHN neurons.
Moreover, the squeezing of lower and upper bounds of the noise interval in which SISR occurs in Fig.\ref{fig:5}\textbf{(b)} and \textbf{(d)}
explains the round-bottom shape of the $CV$-curves (in contrast to the flat-bottom $CV$-curves in Fig.\ref{fig:5}\textbf{(a)} and \textbf{(c)}).

In Fig.\ref{fig:5}, we have a locally ($n_{_1}=1$) coupled ring network topology. 
The numerical simulations (not shown) for  non-locally ($n_{_1}=3,6$) and globally ($n_{_1}=12$) coupled ring network topologies displayed the 
same qualitative behavior observed in Fig.\ref{fig:5} when $\kappa_{_1}$ and $\tau_{_1}$ are varied. 
In Fig.\ref{fig:6}, we fixed $\kappa_{_1}$ and $\tau_{_1}$ and we investigate the effects of 
different ring network topologies on SISR. 

Fig.\ref{fig:6}\textbf{(a)} shows that in a weak ($\kappa_{_1}=0.1$) coupling and short 
($\tau_{_1}=1.0$) time delay regime, changing the ring network topology has no effects on the $CV$-curves which remain 
low with $CV_{min}=0.012$ for $(\sigma_{_1}^{min},\sigma_{_1}^{max})=(4.6\times10^{-7},1.9\times10^{-2})$ for the system of $N=25$ used.
This behavior of SISR is similar to that CR in Fig.\ref{fig:4}\textbf{(a)}
except that the oscillations are less coherent for CR with $CV_{min}=0.14$. The reasons for this behavior is the same as the one 
given for the case of CR Fig.\ref{fig:4}\textbf{(a)}.

For the same reasons given for the case of CR in Fig.\ref{fig:4}\textbf{(b)}, the network topology significantly affects SISR only when the coupling 
and the time delay between the neurons take on larger values. This is observed in Fig.\ref{fig:6}\textbf{(b)} ($\kappa_{_1}=1.0$, $\tau_{_1}=7.0$)
which shows that the sparser the network is, the higher the $CV$-curve is, and hence, the less coherent the oscillations due to SISR are.
Notice further that in Fig.\ref{fig:6}\textbf{(b)}, irrespective of the ring network topology, the $CV$-curves all have a round-bottom shape
due to the shrinking (on both ends) of the noise interval $(\sigma_{_1}^{min},\sigma_{_1}^{max})$, caused by the strong coupling force and the long time delay.

The behavior of the $CV$-curves in  Fig.\ref{fig:6}\textbf{(a)} and \textbf{(b)} can also be explained in terms of the 
interaction potential in Eq.\ref{eq:5b}. When the coupling force becomes weaker $\kappa_{_1}\rightarrow0$ and the time delay shorter $\tau_{_1}\rightarrow0$,
the interaction part of the potential in Eq.\ref{eq:5b} turns to zero (i.e., 
$\frac{\kappa_{_1}}{2n_{_1}}\sum\limits_{j=i-n_{_1}}^{i+n_{_1}}\big[v_{_{1i}}(t)v_{_{1j}}(t-\tau_{_1})-\frac{1}{2}v_{_{1i}}(t)^2\big]\rightarrow0$) and different 
values of $n_{_1}$ (i.e., different ring network topologies) cannot significantly affect this limit and hence cannot also significantly
affect the $CV$-curves as seen in Fig.\ref{fig:6}\textbf{(a)}. Furthermore, in these limits 
(i.e., $\kappa_{_1}\rightarrow0$ and $\tau_{_1}\rightarrow0$), the energy barriers are low (see Fig.\ref{fig:5a}) 
leading to a frequent (and coherent, because the conditions in Eq.\ref{eq:4} are satisfied) spiking, as seen from the low $CV$ values 
in Fig.\ref{fig:6}\textbf{(a)}.

On the other hand, in the strong coupling limit ($\kappa_{_1}\gg0$) and long time delay limit ($\tau_{_1}\gg0$), the interaction part of the 
potential given in Eq.\ref{eq:5b}, that is,
$\frac{\kappa_{_1}}{2n_{_1}}\sum\limits_{j=i-n_{_1}}^{i+n_{_1}}\big[v_{_{1i}}(t)v_{_{1j}}(t-\tau_{_1})-\frac{1}{2}v_{_{1i}}(t)^2\big]$, does not longer turn to zero and
different values of $n_{_1}$ can now  significantly affect the energy barrier functions. 
As $n_{_1}$ decreases (i.e., as the network becomes sparser), the quantify $\frac{\kappa_{_1}}{2n_{_1}}$ (for a fixed $\kappa_{_1}\gg0$)
increases, which means the energy barriers become deeper (see Fig.\ref{fig:5a} with larger values of $\kappa_{_1}$). And as we saw earlier, to be able to jump over deep energy barriers, 
the noise amplitude should be large enough, and large noise amplitudes deteriorate the coherence of the oscillations. This is what we observe 
in Fig.\ref{fig:6}\textbf{(b)}.

Another behavior we can observe is that the $CV$-curves of SISR in Fig.\ref{fig:5} and Fig.\ref{fig:6} are rougher than those 
of CR in Fig.\ref{fig:3} and Fig.\ref{fig:4}, even though the coherence of the oscillations due to SISR is higher than those due to CR.
The reason for this behavior lies in the fact that $CV$-curves are computed using the interspike interval of the fast variable ($v_{_{1i}}$) equations.
Because in the case of SISR, the noise term is attached to fast variable equations, the noise affect the fast variable $v_{_{1i}}$ 
directly and more than the slow variable $w_{_{1i}}$ which is indirectly affected. Hence the roughness of the $CV$-curves. 
While in the case of CR, even though the noise term is instead 
attached to the slow variable ($w_{_{1i}}$) equations, the $CV$-curves are still computed using the interspike interval of the fast variable 
which is in this case, not directly affected by the noise. Hence, we have relatively smoother $CV$-curves with CR than with SISR. 
 
\subsection{Multiplex Network and control of CR}
It has been shown in \cite{Semenova1 and Zakharova 2018} that in a two-layer multiplex network (without time delays),
CR in one layer could induce and enhance CR in the other layer, even with a weak multiplexing. 
Here, we address the question whether a control of CR based on multiplexing is still possible when we use a different noise-induced phenomenon, 
namely, SISR instead of CR. Then, we address a second question: which of the enhancement strategies, CR or SISR,  of CR in one layer of the multiplex network is better?

To answer these questions, we consider layer 1 and layer 2, each with a network size of $N=25$, 
coupled in a multiplex fashion as in Fig.\ref{fig:1a}. 
The coupled layers are non-identical, as we set layer 2 in a locally coupled ring network topology 
with a strong intra-layer coupling force, a long intra-layer time delay, while layer 1 has a globally coupled ring topology
with a weak coupling force and a short time delay. With this choice of parameters,
we set layer 2 in a regime where CR is either poor (i.e., $0.5<CV_{min}<1.0$ for $n_{_2}=1$, $\kappa_{_2}=1.0$, $\tau_{_2}=7.0$, 
see the red curve in Fig.\ref{fig:3}\textbf{(b)}) or non-existent (i.e., $CV_{min}>1.0$ for $n_{_2}=1$, $\kappa_{_2}=1.5$, $\tau_{_2}=7.0$, 
see the pink curve in Fig.\ref{fig:3}\textbf{(d)}). In layer 1, we set the parameters such that CR (see Fig.\ref{fig:4}\textbf{(a)})
or SISR (see Fig.\ref{fig:6}\textbf{(a)}) is very pronounced, i.e., $n_{_1}=12$, $\kappa_{_1}=0.1$, $\tau_{_1}=1.0$.

The multiplexing between the two layers introduces two additional
parameters: the inter-layer time delay $\tau_{_{12}}$ and the inter-layer coupling force $\kappa_{_{12}}$ that characterize the coupling between the layers.
Numerical investigations have shown that the more pronounced CR or SISR in layer 1 is, the more pronounced CR induced in layer 2 would be.
Therefore, to maximize the enhancement or induction of a poor or non-existent CR in layer 2, we use the intra-layer 
control parameters of layer 1 which induce one of the most pronounced forms of CR or SISR
in this layer, i.e., $\kappa_{_1}=0.1$, $\tau_{_1}=0.5$, and $n_{_1}=12$.

\begin{figure}
\begin{center}
\includegraphics[width=4.5cm,height=4.25cm]{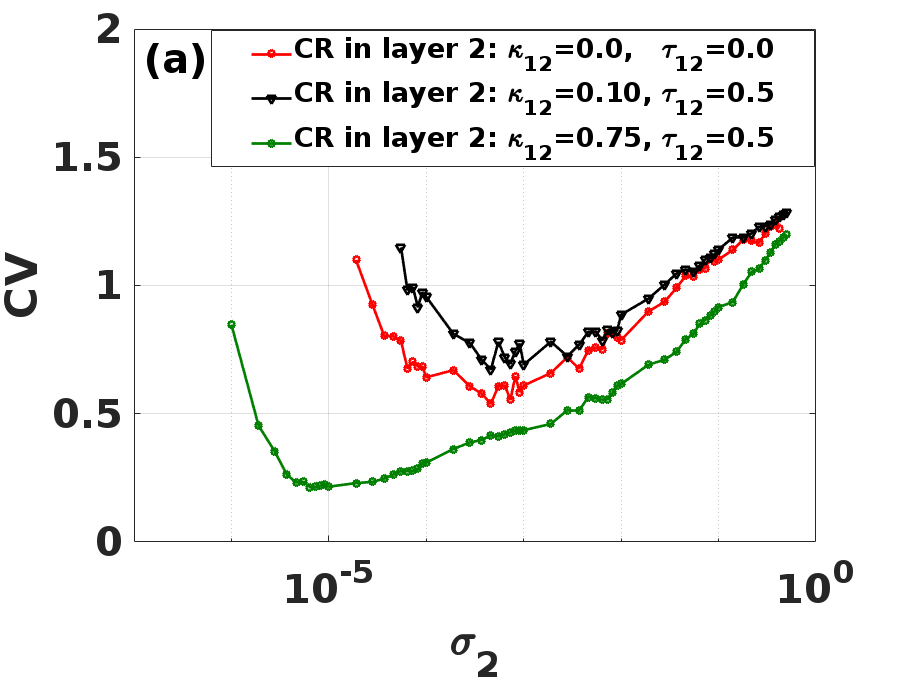}\includegraphics[width=4.5cm,height=4.25cm]{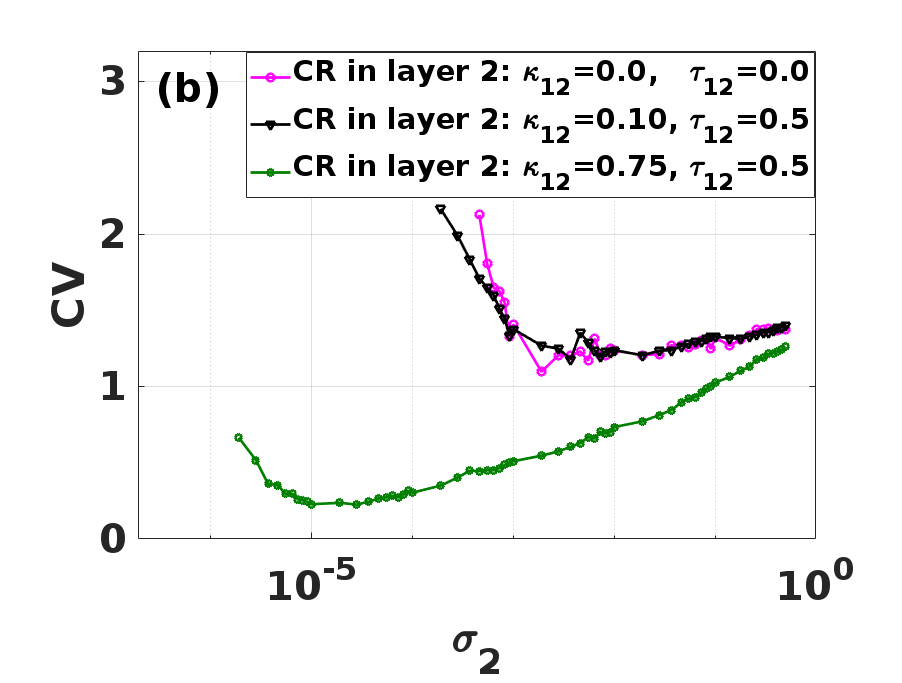}
\caption{(Color online) Coefficient of variation $CV$ against noise amplitude $\sigma_{_2}$ of layer 2 coupled to layer 1 in a multiplexed network.
In \textbf{(a)} and \textbf{(b)} (black curves), when there is a weak inter-layer coupling ($\kappa_{_{12}}=0.10$)
and a short inter-layer time delay ($\tau_{_{12}}=0.5$), a pronounced SISR in layer 1 cannot enhance a poor 
(or non-existent) CR in layer 2. But with a strong inter-layer coupling ($\kappa_{_{12}}=0.75$) (green curves in \textbf{(a)} and \textbf{(b)}) and 
a short inter-layer time delay ($\tau_{_{12}}=0.5$), a pronounced SISR in layer 1 can enhance a poor or non-existent CR in layer 1.
Parameters of layer 1: $\varepsilon_{_1}=0.0005$, $\kappa_{_1}=0.1$, $\tau_{_1}=0.5$, $n_{_1}=12$, $\sigma_{_1}=\sigma_{_2}$, $\beta_{_1}=0.75$, $\alpha=0.5$, $N=25$.
Parameters of layer 2: $\varepsilon_{_2}=0.01$, $\tau_{_2}=7.0$, $n_{_2}=1$, $\beta_{_2}=0.75$, $\alpha=0.5$, $N=25$,  $\kappa_{_2}=1.0$ in \textbf{(a)}, 
$\kappa_{_2}=1.5$ in \textbf{(b)}.
\label{fig:7}}
\end{center}
\end{figure}

We now investigate the impacts of the inter-layer coupling force $\kappa_{_{12}}$, 
the inter-layer time delay $\tau_{_{12}}$, and a pronounced SISR in layer 1 on a poor or non-existent CR in layer 2.
In Fig.\ref{fig:7}\textbf{(a)}, we see that at a weak ($\kappa_{_{12}}=0.10$) inter-layer coupling force and a 
short ({$\tau_{_{12}}=0.5$}) inter-layer time delay, the poor CR in layer 2 
cannot be enhanced by a pronounced SISR in layer 1. 
In isolation ($\kappa_{_{12}}$=0.0), layer 2 has $CV_{min}=0.56$ at $\sigma_{_2}=4.6\times10^{-4}$ (see red curve in Fig.\ref{fig:7}\textbf{(a)}), 
and for a weak inter-layer coupling force ($\kappa_{_{12}}=0.10$), layer 2 shows a $CV$-curve with a rather higher minimum value, i.e.,
$CV_{min}=0.67$ at $\sigma_{_2}=4.6\times10^{-4}$ (see black curve in Fig.\ref{fig:7}\textbf{(a)}).  

This inability of SISR to enhance CR in a weak inter-layer coupling regime is also observed in Fig.\ref{fig:7}\textbf{(b)}, where CR is non-existent. 
Here, layer 1 in isolation ($\kappa_{_{12}}=0.0$) shows a $CV$-curve with 
$CV_{min}=1.14$ at $\sigma_{_2}=6.4\times10^{-3}$ (see pink curve) 
against $CV_{min}=1.18$ at $\sigma_{_2}=3.7\times10^{-3}$ (see black curve) with a weak coupling force of $\kappa_{_{12}}=0.10$ and short time delay
of $\tau_{_{12}}=0.5$. Therefore, at a weak inter-layer coupling force, a pronounced SISR in layer 1, instead of improving CR in layer 2, can actually make CR even poorer than when layer 2 is isolated.

On the other hand, strong inter-layer coupling forces drastically reverse the scenario. 
From the green curves in Fig.\ref{fig:7}\textbf{(a)} and \textbf{(b)},
the inter-layer coupling force takes a larger value, i.e., $\kappa_{_{12}}=0.75$ and the time delay is kept fixed at $\tau_{_{12}}=0.5$. We observe that 
the $CV$-curves of layer 2 drop significantly to low values, indicating an enhancement of a poor CR and an induction of a non-existent CR in layer 2 
by a pronounced SISR in layer 1.
In Fig.\ref{fig:7}\textbf{(a)}, the relatively high values of the minima of the $CV$-curves of layer 2 
(i.e., $CV_{min}=0.56$ with no ($\kappa_{_{12}}=0.0$) multiplexing and $CV_{min}=0.67$ with a weak ($\kappa_{_{12}}=0.10$) multiplexing) are shifted 
to a low value of $CV_{min}=0.21$ at $\sigma_{_2}=6.4\times10^{-6}$, indicating a high coherence of the oscillations in layer 2 induced
by a strong inter-layer coupling force and a pronounced SISR in layer 1.
We can also see in Fig.\ref{fig:7}\textbf{(b)} that the very high values of the minima of the $CV$-curves of layer 2 (i.e., $CV_{min}=1.14$ 
with $\kappa_{_{12}}=0.0$ and $CV_{min}=1.18$ with $\kappa_{_{12}}=0.10$) are shifted to a significantly low value, i.e., $CV_{min}=0.23$ 
at $\sigma_{_2}=2.8\times10^{-5}$ by a strong inter-layer coupling force and a pronounced SISR in layer 1.

Now, in the last part of our study, we compare the performance of SISR (in layer 1) in enhancing CR in layer 2 to the performance
of CR (in layer 1) in enhancing CR in layer 2.
In the control of CR by SISR (which we henceforth refer to as SISR-CR control scheme), we set layer 1 with a pronounced SISR, i.e., we choose a globally coupled ring network topology 
($n_{_1}=12$), a weak intra-layer coupling force ($\kappa_{_1}=0.1$), and
short intra-layer time delay ($\tau_{_1}=0.1$), with $\sigma_{_1}\neq0$ and $\sigma_{_3}=0$.

In the control of CR by CR (which we henceforth refer as to CR-CR control scheme), we set layer 1 such that it also has a pronounced CR with of course,
the same ring network topology, the same intra-layer coupling force, and the same intra-layer time delay as in the SISR-CR control scheme, i.e., 
we choose a globally coupled ring network topology ($n_{_1}=12$), a weak intra-layer coupling force ($\kappa_{_1}=0.1$) and a short intra-layer time delay ($\tau_{_1}=0.1$).
Then, we switch off the noise term ($\sigma_{_1}=0$) 
on the fast variable ($v_{_{1i}}$) equations in layer 1 and switch on the noise term ($\sigma_{_3}\neq0$) on the slow variable ($w_{_{1i}}$) equations of this layer.
We recall that for the FHN neuron model, SISR requires the noise term to be attached to the fast variable equation, while CR requires the noise 
term to be attached to the slow variable equation.
Furthermore, because CR (unlike SISR with $\varepsilon_{_1}=0.0005$) does not necessarily require a singular limit 
($\varepsilon_{_2}\rightarrow0$) to occur, we set the singular parameter $\varepsilon_{_2}$ 
to its standard value $\varepsilon_{_2}=0.01\ll\varepsilon_{_1}$. We further note that the coherence of the noise-induced
oscillations due to CR is insensitive against variations of the timescale
separation ratio between the fast and slow variables of the excitable system \cite{Muratov et al 2005,Lee DeVille et al. 2005}.
 
In order to control CR in layer 2, we choose an intra-layer coupling force, intra-layer 
time delay, and a ring network topology such that CR is non-existent in this layer when it is in isolation, 
i.e., we choose $\kappa_{_2}=1.5$, $\tau_{_2}=7.0$, $n_{_2}=1$, see in Fig.\ref{fig:3}\textbf{(d)}.
In the SISR-CR and CR-CR control schemes, layer 1 and layer 2 are coupled in a multiplex fashion 
in two different inter-layer coupling regimes. In the first regime, we choose weak 
inter-layer coupling forces ($\kappa_{_{12}}=0.2$ and $\kappa_{_{12}}=0.4$) and a short inter-layer time delay ($\tau_{_{12}}=1.0$).
In the second regime, we choose strong inter-coupling forces ($\kappa_{_{12}}=0.5$ and $\kappa_{_{12}}=1.0$) with the same short time delay ($\tau_{_{12}}=1.0$). 
We compare the performances of SISR-CR and CR-CR control schemes in enhancing CR in layer 2 in the weak and strong coupling regimes. 

\begin{figure}
\begin{center}
\includegraphics[width=4.5cm,height=4.25cm]{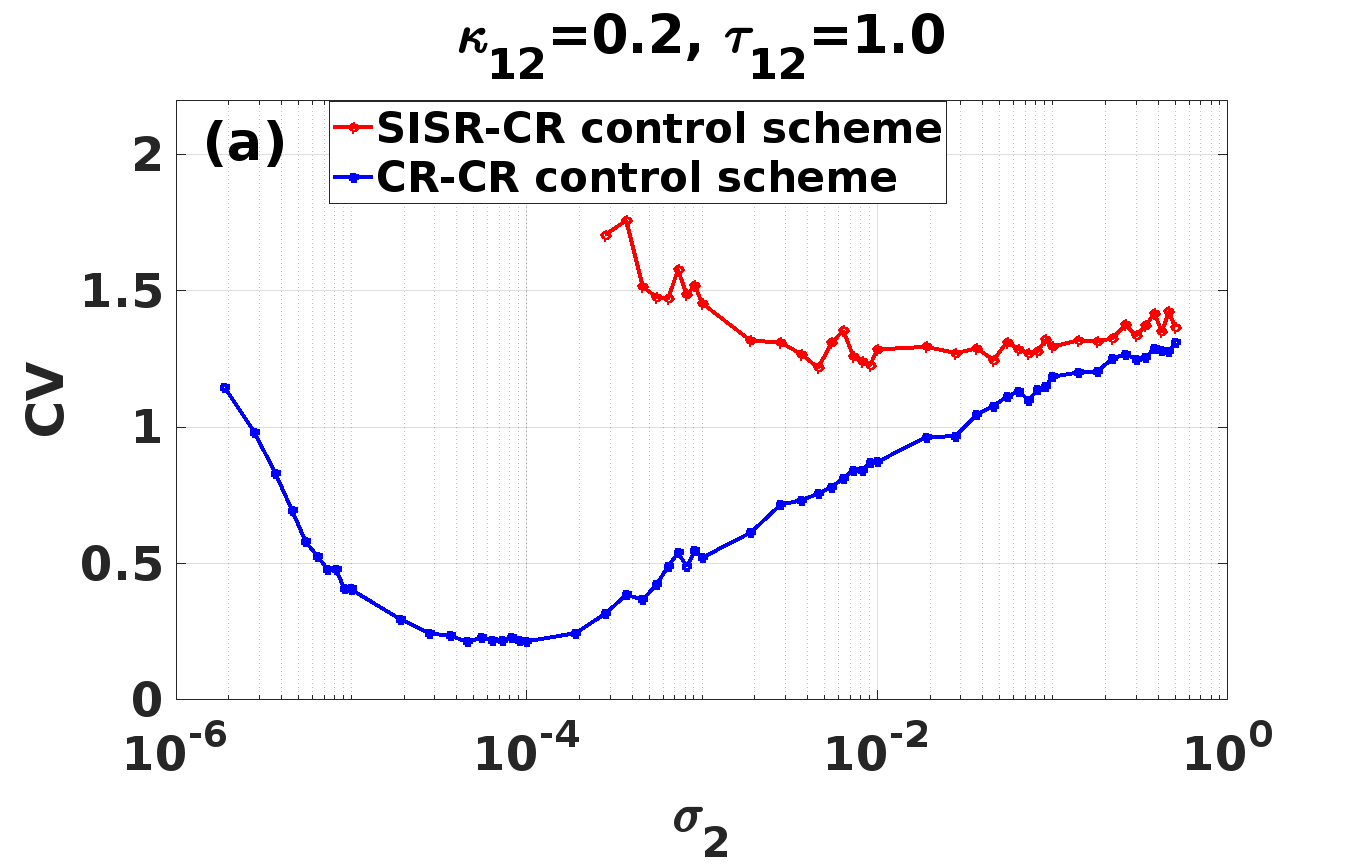}\includegraphics[width=4.5cm,height=4.25cm]{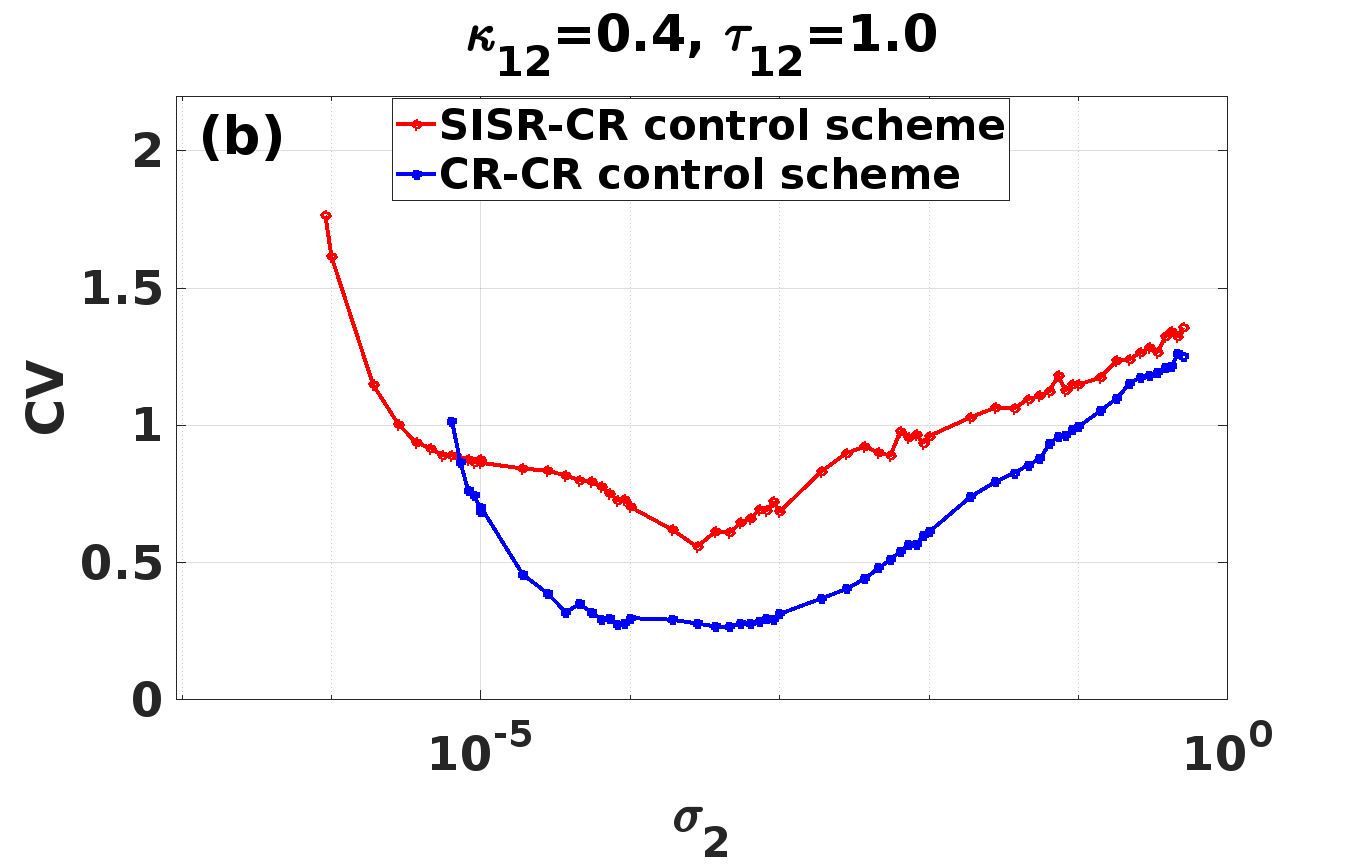}
\includegraphics[width=4.5cm,height=4.25cm]{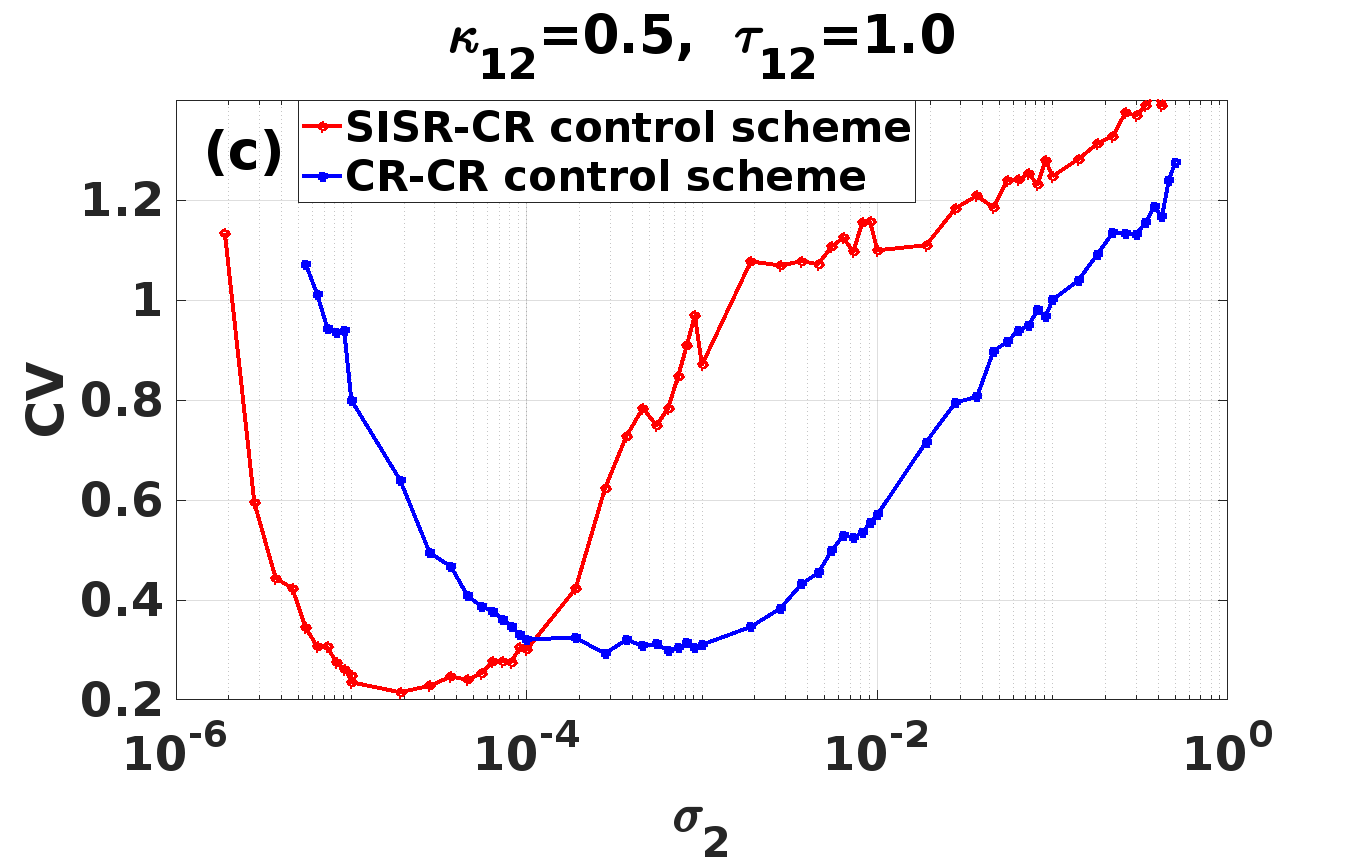}\includegraphics[width=4.5cm,height=4.25cm]{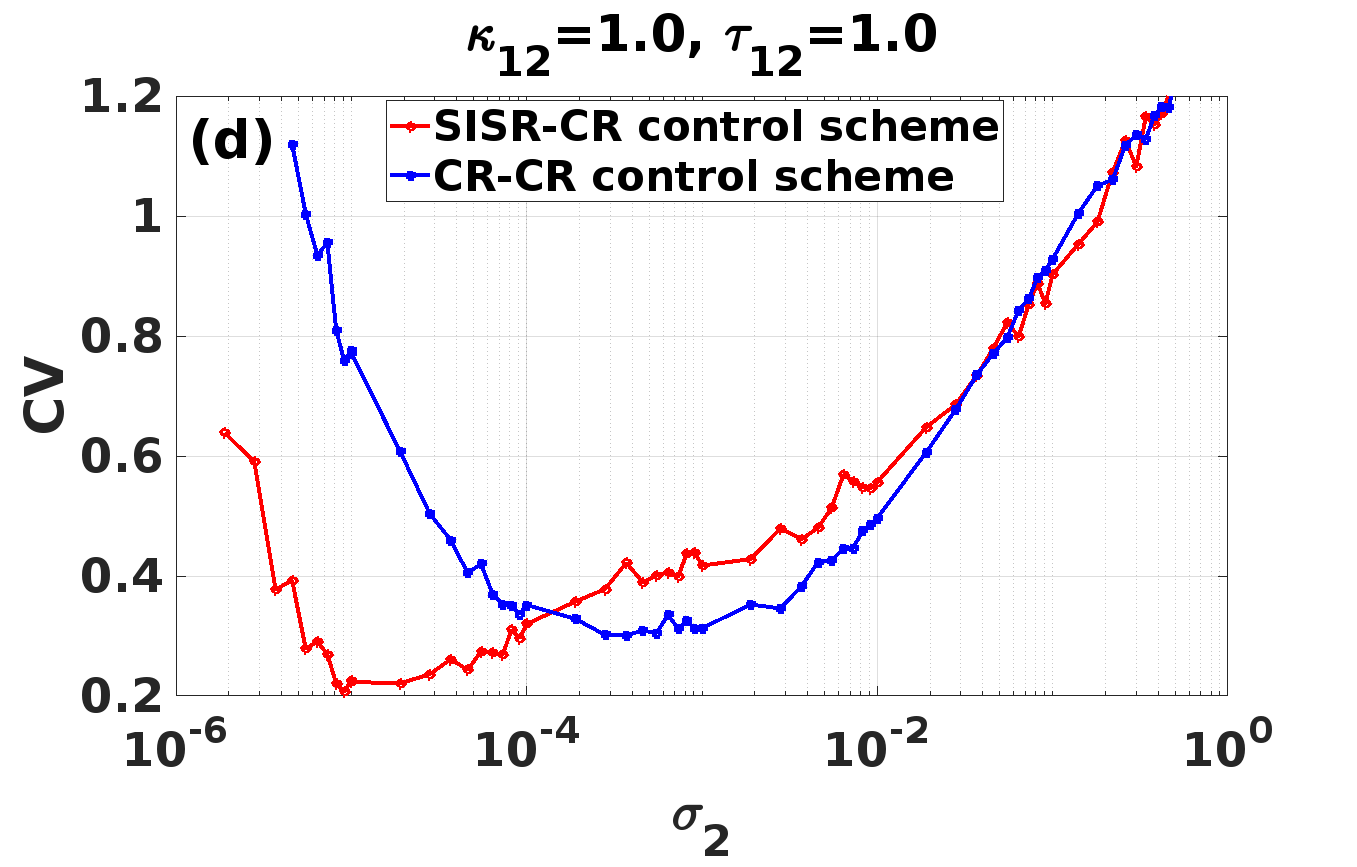}
\caption{(Color online) Coefficient of variation $CV$ against noise amplitude $\sigma_{_2}$ of layer 2 coupled to layer 1 in a multiplexed network.
In \textbf{(a)} and \textbf{(b)}, we have a weak inter-layer coupling regime, where the CR-CR control scheme performs better than the SISR-CR control scheme in enhancing CR in layer 2 of the multiplex. 
In \textbf{(c)} and \textbf{(d)}, we have a strong inter-layer coupling regime, where the SISR-CR control scheme takes over and performs better than the CR-CR control scheme. 
Parameters of layer 1 with CR: $\varepsilon_{_1}=0.01$, $\kappa_{_1}=0.1$, $\tau_{_1}=0.1$, $n_{_1}=12$, $\sigma_{_3}=\sigma_{_2}$, $\beta_{_1}=0.75$, $\alpha=0.5$, $N=25$.
Parameters of layer 1 with SISR: $\varepsilon_{_1}=0.0005$, $\kappa_{_1}=0.1$, $\tau_{_1}=0.1$, $n_{_1}=12$, $\sigma_{_1}=\sigma_{_2}$, $\beta_{_1}=0.75$, $\alpha=0.5$, $N=25$.
Parameters of layer 2: $\varepsilon_{_2}=0.01$,  $\kappa_{_2}=1.5$, $\tau_{_2}=7.0$, $n_{_2}=1$, $\beta_{_2}=0.75$, $\alpha=0.5$, $N=25$.\label{fig:8}}
\end{center}
\end{figure}

In Fig.\ref{fig:8}\textbf{(a)} with $\kappa_{_{12}}=0.2$ and $\tau_{_{12}}=1.0$, we see that the CR-CR control scheme (blue curve)
performs much better than the SISR-CR control scheme (red curve). Here, the CR-CR control scheme provides a minimum $CV$ of
$CV_{min}=0.21$ at $\sigma_{_2}=4.6\times10^{-5}$ against $CV_{min}= 1.22$ at  $\sigma_{_2}=4.6\times10^{-3}$
for the SISR-CR control scheme. This ability of CR (in one layer of a multiplex network) to enhance CR (in the other layer) in a weak inter-layer coupling
regime was already shown in \cite{Semenova1 and Zakharova 2018}. Our work further confirms their finding on a different version of the FHN model.
This behavior is however, absent with SISR which performs very poorly in the SISR-CR control scheme in a weak inter-layer coupling regime.

In Fig.\ref{fig:8}\textbf{(b)}, the inter-layer coupling force is increased to $\kappa_{_{12}}=0.4$ and the short inter-layer time delay is kept fixed at 
$\tau_{_{12}}=1.0$. The SISR-CR control scheme improves on its performance which is, however, still poorer than that of the CR-CR control scheme. 
Here, the SISR-CR control scheme provides a minimum $CV$ of $CV_{min}= 0.56$ at $\sigma_{_2}=2.8\times10^{-4}$ against  $CV_{min}=0.23$ at $\sigma_{_2}=4.6\times10^{-4}$ 
for the CR-CR control scheme.

Surprisingly, in Fig.\ref{fig:8}\textbf{(c)}, where we have a strong inter-layer coupling force ($\kappa_{_{12}}=0.5$) 
and the same short inter-layer time delay ($\tau_{_{12}}=1.0$),
the SISR-CR control scheme takes over and performs better than the CR-CR control scheme, especially at weaker noise amplitudes.
Here, the SISR-CR control scheme provides a minimum $CV$ of $CV_{min}=0.22$ at $\sigma_{_2}=1.9\times10^{-5}$ against 
$CV_{min}=0.29$ at $\sigma_{_2}=2.8\times10^{-4}$ for the CR-CR control scheme.

In Fig.\ref{fig:8}\textbf{(d)}, we further increase the strength of the inter-layer coupling force to $\kappa_{_{12}}=1.0$ 
and the SISR-CR control scheme performs better than when we have $\kappa_{_{12}}=0.5$ and is still better than the CR-CR control scheme. 
Here, the SISR-CR control scheme provides minimum $CV$ of $CV_{min}=0.20$ at $\sigma_{_2}=9.1\times10^{-6}$ against $CV_{min}=0.30$ 
at $\sigma_{_2}=5.5\times10^{-4}$ for the CR-CR control scheme.

\begin{figure}
\begin{center}
\includegraphics[width=4.5cm,height=4.25cm]{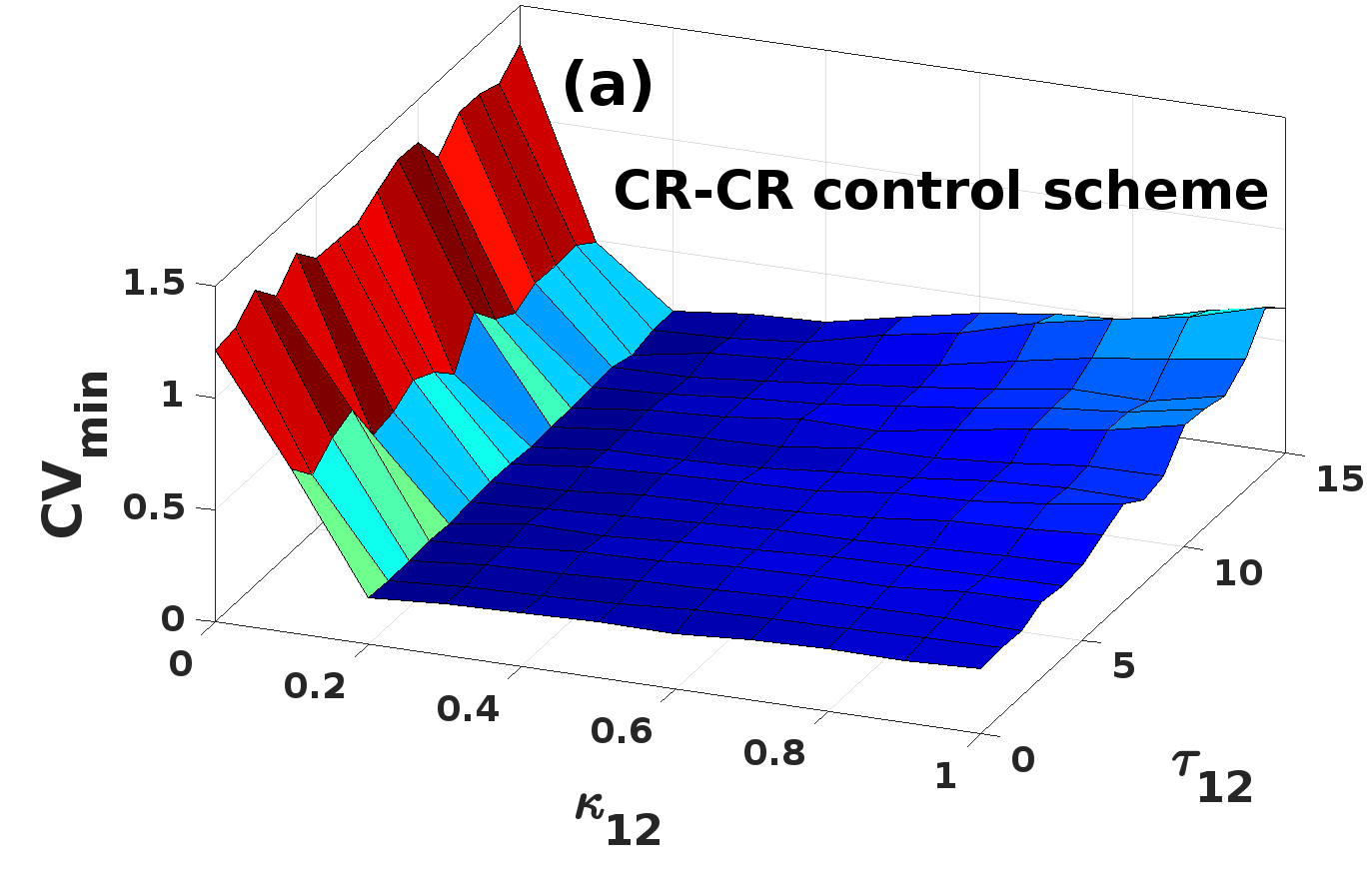}\includegraphics[width=4.5cm,height=4.25cm]{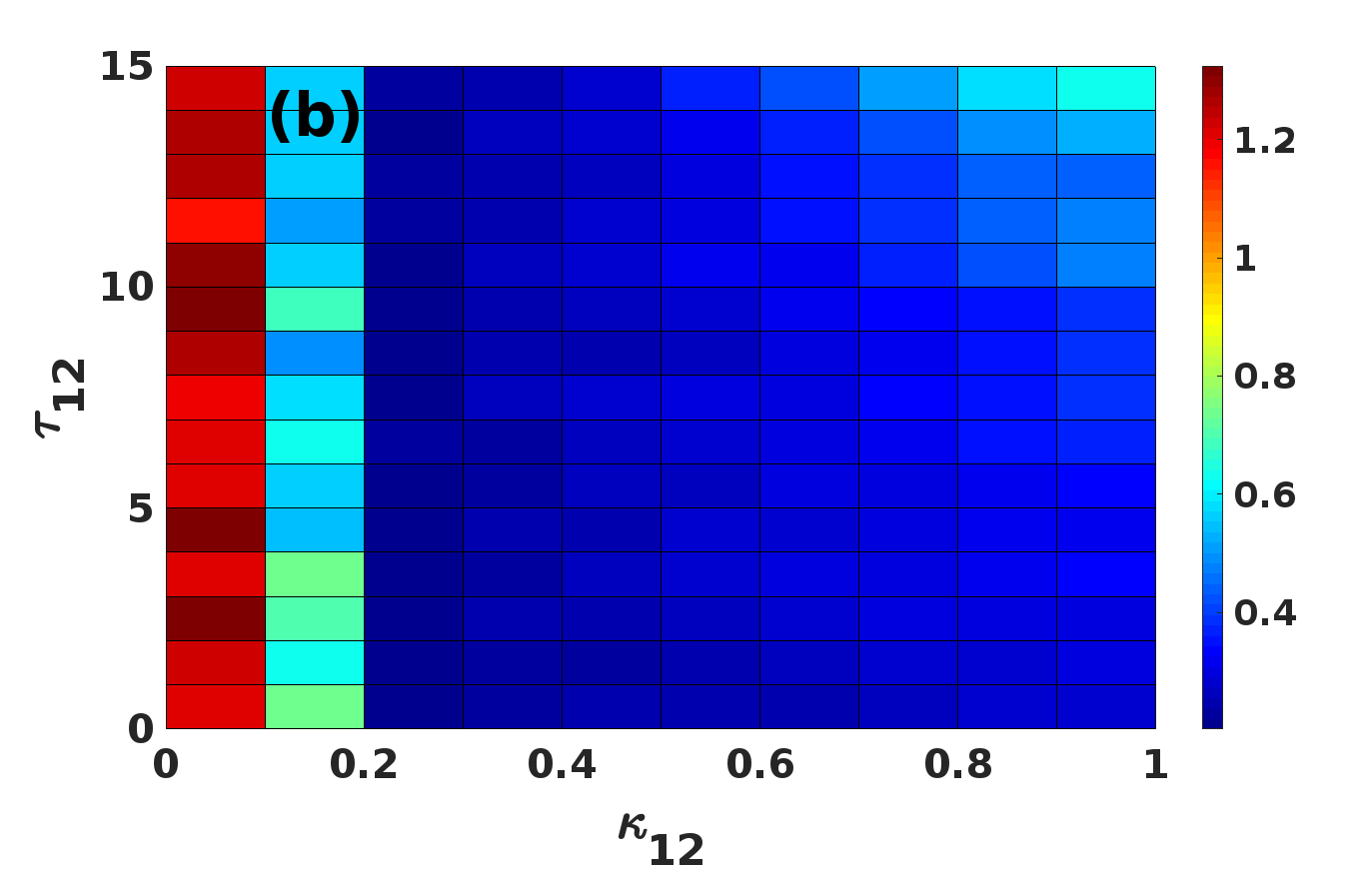}
\includegraphics[width=4.5cm,height=4.25cm]{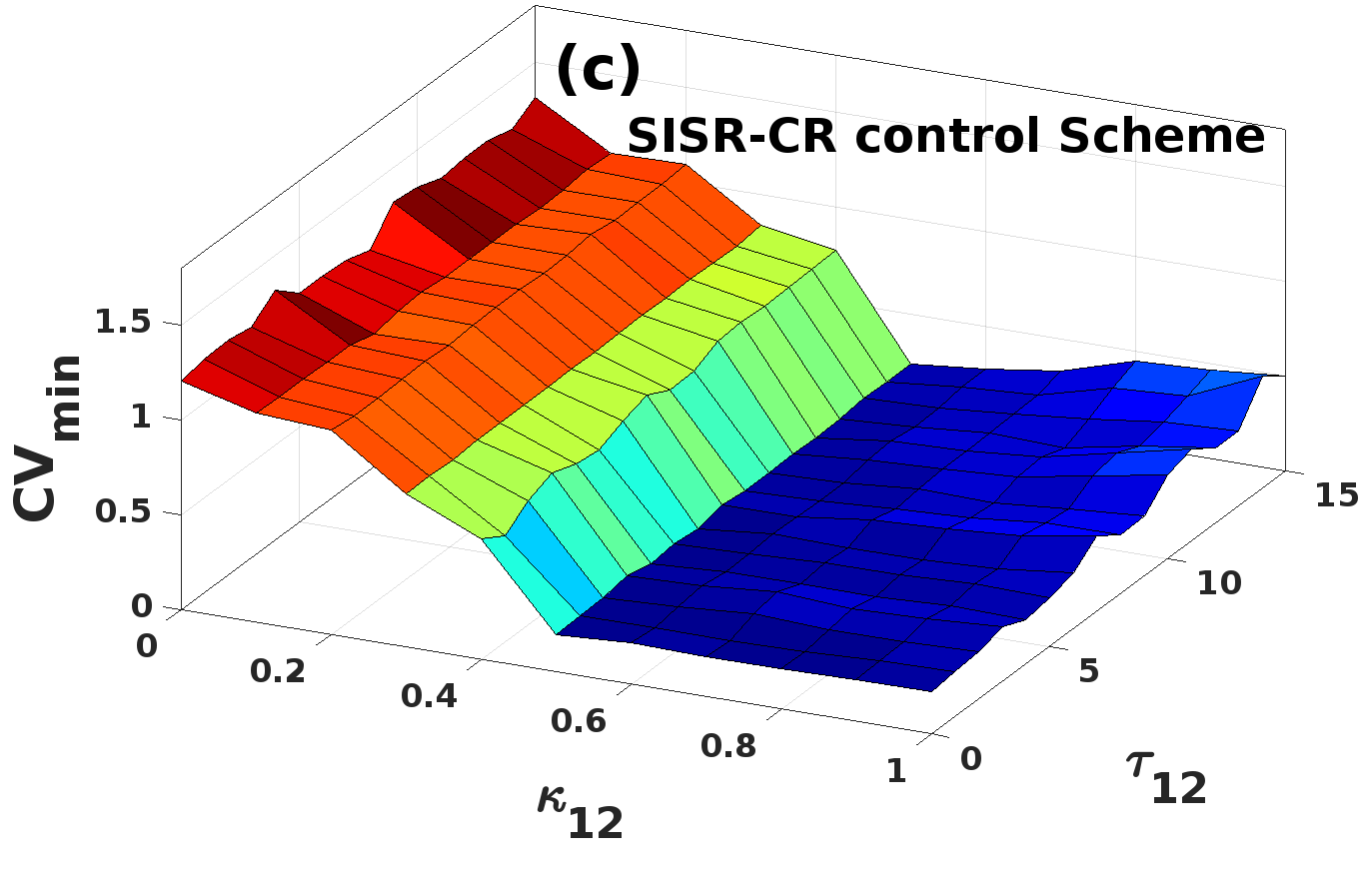}\includegraphics[width=4.5cm,height=4.25cm]{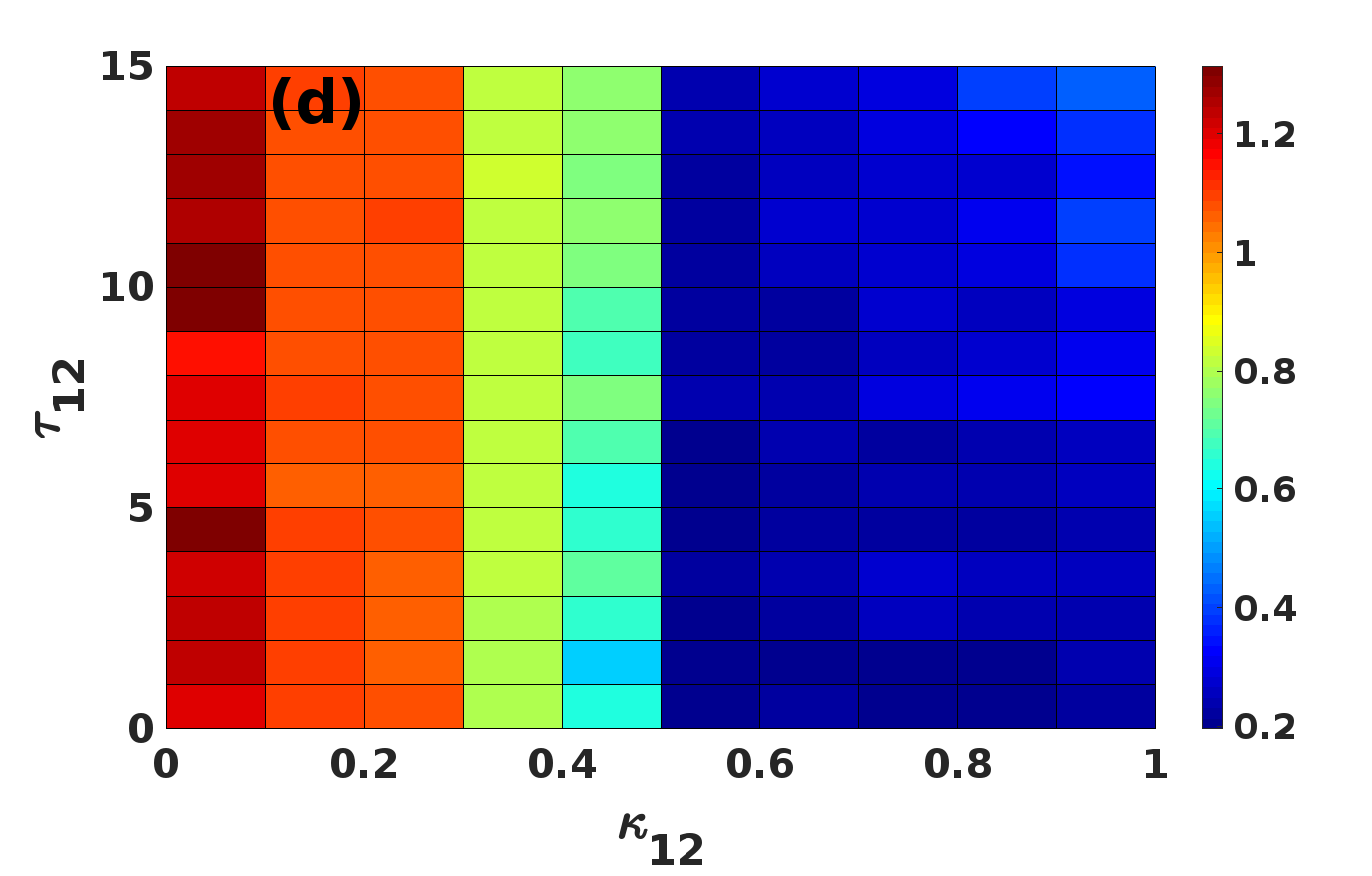}
\caption{(Color online) Minimum coefficient of variation ($CV_{min}$) against the inter-layer coupling force $\kappa_{_{12}}$ and time delay $\tau_{_{12}}$ 
for layer 2 when coupled to layer 1 in a multiplexed network.
In \textbf{(a)} and \textbf{(b)}, we have a 3D plot and its 2D projection onto the parameter plane 
($\kappa_{_{12}}-\tau_{_{12}}$) showing the enhancement performance of the CR-CR control scheme, which is good even 
at weak ($0.1<\kappa_{_{12}}<0.5$) inter-layer coupling forces. In \textbf{(c)} and \textbf{(d)}, we have a 3D plot and its 2D projection
onto the parameter space ($\kappa_{_{12}}-\tau_{_{12}}$) showing the enhancement performance of the SISR-CR control 
scheme, which is poor at weak ($0.0\leq\kappa_{_{12}}<0.5$) inter-layer coupling forces. But at strong ($0.5\leq\kappa_{_{12}}\leq1.0$) 
inter-layer coupling forces, the SISR-CR control 
scheme has a good performance which is even better than that of the CR-CR control scheme, especially at longer inter-layer time delays and weaker noise amplitudes.  
Parameters of layer 1 with CR: $\varepsilon_{_1}=0.01$, $\kappa_{_1}=0.1$, $\tau_{_1}=0.1$, $n_{_1}=12$, $\sigma_{_3}=\sigma_{_2}$, $\beta_{_1}=0.75$, $\alpha=0.5$, $N=25$.
Parameters of layer 1 with SISR: $\varepsilon_{_1}=0.0005$, $\kappa_{_1}=0.1$, $\tau_{_1}=0.1$, $n_{_1}=12$, $\sigma_{_1}=\sigma_{_2}$, $\beta_{_1}=0.75$, $\alpha=0.5$, $N=25$.
Parameters of layer 2: $\varepsilon_{_2}=0.01$,  $\kappa_{_2}=1.5$, $\tau_{_2}=7.0$, $n_{_2}=1$, $\beta_{_2}=0.75$, $\alpha=0.5$, $N=25$.\label{fig:9}}
\end{center}
\end{figure}

In Fig.\ref{fig:9}, we globally compare the performances of the SISR-CR and CR-CR control schemes in the $(\kappa_{_{12}}-\tau_{_{12}})$ parameter space.
In Fig.\ref{fig:9}\textbf{(a)} and \textbf{(b)}, we have the CR-CR control scheme with a color coded minimum $CV$ ($CV_{min}$). We can see that for
\textit{very} weak inter-layer coupling regime ($0\leq\kappa_{_{12}}<0.2$) the CR-CR control scheme performs poorly as $CV_{min}$ is very high in value (red color). 
But as soon as $\kappa_{_{12}}\geq0.2$, $CV_{min}$ drops to dark blue indicating a good performance of the CR-CR control scheme at weak 
inter-layer coupling forces. 
Furthermore, we observe that as the inter-layer time delay increases, $CV_{min}$ also increases. 
This effect of the inter-layer time delay is however more significant in stronger inter-layer coupling regimes i.e., $\kappa_{_{12}}\geq0.7$.

In Fig.\ref{fig:9}\textbf{(c)} and \textbf{(d)} we have the SISR-CR control scheme. 
We can see that for weak inter-layer coupling regime ($0\leq\kappa_{_{12}}<0.5$), the SISR-CR control scheme 
performs poorly as the color coded $CV_{min}$ takes red and green colors indicating 
a relatively higher $CV_{min}$ compared to those of the CR-CR control scheme. 
However, in the strong inter-layer coupling regime ($0.5\leq\kappa_{_{12}}\leq1.0$), 
the color coded $CV_{min}$ takes a dark blue (darker than the blue color in the 
CR-CR control scheme, especially at long time delays) 
color, indicating a high (higher than those of the CR-CR control scheme) enhancing 
performance of the SISR-CR control scheme. Moreover, we can also observe (like in the CR-CR control scheme) 
that, as the inter-layer time delay increases, the performance of the SISR-CR control scheme reduces 
with the effect becoming more significant at stronger ($\kappa_{_{12}}\geq0.8$) inter-layer coupling forces. 
This was also observed in the isolated layers, where the intra-layer time delay 
significantly affected the coherence of the oscillations only in a strong intra-layer coupling regime, 
see Fig.\ref{fig:3}\textbf{(a)}-\textbf{(b)} and Fig.\ref{fig:5}\textbf{(a)}-\textbf{(b)} .

\section{Summary and Conclusions} \label{section4}
In this paper, we have considered a two-layer time-delayed multiplex network of stochastic FHN neurons in the excitable regime.
First, we systematically and independently investigated the impacts of the intra-layer time-delayed couplings and different ring network topologies
on the noise-induced phenomena of CR and SISR in the isolated layers of the multiplex network.
We determined the parameter regimes and the network topology which enhance or weaken CR and SISR the most. 
It was found that strong intra-layer coupling forces, long intra-layer time delays, and a locally coupled ring network topology tend to 
weaken CR and SISR, with CR showing a more much sensitive behavior to variations of these control parameters than SISR.
Furthermore, in the isolated layer network of FHN neurons exhibiting SISR, we found that the maximum noise amplitude at which SISR 
occurs is not fixed, but controllable, especially in a strong intra-layer coupling force and long intra-layer time delay regime.
This is in contrast to SISR in a single isolated FHN neuron, where this maximum noise amplitude is fixed and uncontrollable.

Secondly, we set up two control strategies of CR in layer 2, based not only on the multiplexing with layer 1, but also on SISR and CR.
In one of the control schemes (which we termed as CR-CR control scheme), we have a pronounced ($n_{_1}=12$, $\kappa_{_1}=0.1$, $\tau_{_1}=0.5$) 
CR in layer 1 which controls CR in layer 2.
While in the second control scheme (SISR-CR control scheme), we have a pronounced ($n_{_1}=12$, $\kappa_{_1}=0.1$, $\tau_{_1}=0.5$)
SISR in layer 1 which controls CR in layer 2.
In our control schemes, we have considered layer 2 of the multiplex network with a non-optimal network topology and time-delayed coupling for CR, 
i.e., the intra-layer time-delayed coupling of layer 2 was such that CR is non-existent when the layer is 
in isolation (i.e., $n_{_2}=1$, $\kappa_{_2}=1.5$, $\tau_{_2}=7.0$).

We found that the multiplexing of layers even with a weak (but not too weak) inter-layer coupling force, that is $0.2\leq\kappa_{_{12}}<0.5$, 
the CR-CR control scheme could significantly enhance (after inducing a previously non-existent) 
CR in layer 2. While in the weak inter-layer coupling regime ($0.0\leq\kappa_{_{12}}<0.5$), 
the SISR-CR control scheme could not significantly enhance CR, nor even, at very weak ($0.0\leq\kappa_{_{12}}<0.3$) coupling forces, induce CR. 
But at strong inter-layer coupling force ($\kappa_{_{12}}\geq0.5$), the SISR-CR control scheme performs very well in inducing and enhancing 
CR in layer 2. Surprisingly, the SISR-CR control scheme is found to perform better than the CR-CR control scheme in this strong coupling regime, especially 
at weaker noise amplitudes and longer inter-layer time delays. 

Moreover, we found that longer inter-layer time delays inhibit the 
induction and enhancement performances of the CR-CR and SISR-CR control schemes. This inhibition however became more pronounced at 
stronger inter-layer coupling forces, with the CR-CR control scheme suffering the most from this inhibition.
We expect our results to find applications in experimental neuroscience and engineering.

\section*{Acknowledgments}

Marius E. Yamakou thanks the Max Planck Institute for Mathematics in the Sciences, Leipzig, Germany.

\end{document}